\documentclass[twocolumn]{aastex63} 
\usepackage[normalem]{ulem}
\usepackage{chemformula}
\usepackage{hyperref}
\graphicspath{{./}{figures/}}

\shorttitle{A molecular dichotomy in disk dust cavities }

\begin{document}

\title{Protoplanetary disk cavities with JWST-MIRI: a dichotomy in molecular emission}

\author{Patrick Mallaney}
\altaffiliation{These authors contributed proportionately to this work.}
\affil{Department of Physics, Texas State University, 749 N Comanche Street, San Marcos, TX 78666, USA}
\email{pmallane@nd.edu} 

\author[0000-0003-4335-0900]{Andrea Banzatti}
\altaffiliation{These authors contributed proportionately to this work.}
\affil{Department of Physics, Texas State University, 749 N Comanche Street, San Marcos, TX 78666, USA}
\email{banzatti@txstate.edu} 

\author[0000-0003-3682-6632]{Colette Salyk}
\affil{Department of Physics and Astronomy, Vassar College, 124 Raymond Avenue, Poughkeepsie, NY 12604, USA}

\author[0000-0001-7962-1683]{Ilaria Pascucci}
\affil{Department of Planetary Sciences, University of Arizona, 1629 East University Boulevard, Tucson, AZ 85721, USA}

\author[0000-0001-8764-1780]{Paola Pinilla}
\affiliation{Mullard Space Science Laboratory, University College London, Holmbury St Mary, Dorking, Surrey RH5 6NT, UK}

\author[0000-0002-5758-150X]{Joan Najita}
\affiliation{NSF’s NOIRLab, 950 N. Cherry Avenue, Tucson, AZ 85719, USA}

\author[0000-0001-7552-1562]{Klaus M. Pontoppidan}
\affiliation{Jet Propulsion Laboratory, California Institute of Technology, 4800 Oak Grove Drive, Pasadena, CA 91109, USA}

\author[0000-0002-3291-6887]{Sebastiaan Krijt}
\affiliation{School of Physics and Astronomy, University of Exeter, Stocker Road, Exeter EX4 4QL, UK}

\author[0000-0003-0787-1610]{Geoffrey A. Blake}
\affiliation{Division of Geological \& Planetary Sciences, MC 150-21, California Institute of Technology, Pasadena, CA 91125, USA}

\author[0000-0002-1103-3225]{Benoît Tabone}
\affiliation{Université Paris-Saclay, CNRS, Institut d’Astrophysique Spatiale, 91405 Orsay, France}

\author[0000-0001-8240-978X]{Till Kaeufer}
\affiliation{School of Physics and Astronomy, University of Exeter, Stocker Road, Exeter EX4 4QL, UK}

\author[0000-0002-0661-7517]{Ke Zhang}
\affil{Department of Astronomy, University of Wisconsin-Madison, Madison, WI 53706, USA}

\author[0000-0002-7607-719X]{Feng Long}
\altaffiliation{NASA Hubble Fellowship Program Sagan Fellow}
\affil{Department of Planetary Sciences, University of Arizona, 1629 East University Boulevard, Tucson, AZ 85721, USA}

\author[0000-0001-6947-6072]{Jane Huang}
\affiliation{Department of Astronomy, Columbia University, 538 W. 120th Street, Pupin Hall, New York, NY 10027, USA}

\author[0000-0003-4853-5736]{Giovanni Rosotti}
\affiliation{Dipartimento di Fisica, Università degli Studi di Milano, via Giovanni Celoria 16, 20133, Milano, Italy}

\author[0000-0001-8798-1347]{Karin I. \"Oberg}
\affiliation{Center for Astrophysics, Harvard \& Smithsonian, 60 Garden St., Cambridge, MA 02138, USA}

\author[0000-0002-5296-6232]{Mar\'{i}a Jos\'{e} Colmenares}
\affiliation{Department of Astronomy, University of Michigan, Ann Arbor, MI 48109, USA}

\author{Andrew Lay}
\affil{Department of Physics, Texas State University, 749 N Comanche Street, San Marcos, TX 78666, USA}

\author[0000-0002-2828-1153]{Lucas A. Cieza}
\affil{N\'ucleo de Astronom\'ia, Facultad de Ingenier\'ia y Ciencias, Universidad Diego Portales, Av Ej\'ercito 441, Santiago, Chile}

\author[0000-0003-2076-8001]{L. Ilsedore Cleeves}
\affil{Astronomy Department, University of Virginia, Charlottesville, VA 22904, USA}

\author[0009-0008-8176-1974]{Joe Williams}
\affiliation{School of Physics and Astronomy, University of Exeter, Stocker Road, Exeter EX4 4QL, UK}

\author[0000-0001-8184-5547]{Chengyan Xie}
\affil{Department of Planetary Sciences, University of Arizona, 1629 East University Boulevard, Tucson, AZ 85721, USA}

\author[0000-0002-4147-3846]{Miguel Vioque}
\affiliation{European Southern Observatory, Karl-Schwarzschild-Str. 2, 85748 Garching bei München, Germany}

\author[0000-0002-0554-1151]{Mayank Narang}
\affiliation{Jet Propulsion Laboratory, California Institute of Technology, 4800 Oak Grove Drive, Pasadena, CA 91109, USA}

\author[0000-0002-4276-3730]{Nicholas P. Ballering}
\affiliation{Space Science Institute, Boulder, CO 80301, USA}

\author[0000-0001-6218-2004]{Minjae Kim}
\affiliation{Mullard Space Science Laboratory, University College London, Holmbury St Mary, Dorking, Surrey RH5 6NT, UK}

\author{the JDISCS Collaboration}

\begin{abstract}
The evolution of planet-forming regions in protoplanetary disks is of fundamental importance to understanding planet formation. Disks with a central deficit in dust emission, a ``cavity", have long attracted interest as potential evidence for advanced disk clearing by protoplanets and/or winds. Before JWST, infrared spectra showed that these disks typically lack the strong molecular emission observed in full disks. In this work, we combine a sample of 12 disks with millimeter cavities of a range of sizes ($\sim2$--70~au) and different levels of millimeter and infrared continuum deficits. We analyze their molecular spectra as observed with MIRI on JWST, homogeneously reduced with the new JDISCS pipeline. This analysis demonstrates a stark dichotomy in molecular emission where ``molecule-rich" (MR) cavities follow global trends between water, CO, and OH luminosity and accretion luminosity as in full disks, while ``molecule-poor" (MP) cavities are significantly sub-luminous in all molecules except sometimes OH. Disk cavities generally show sub-luminous organic emission, higher OH/\ch{H2O} ratios, and suggest a lower water column density. The sub-thermal excitation of CO and water vibrational lines suggests a decreased gas density in the emitting layer in all cavities, supporting model expectations for \ch{C2H2} photodissociation. We discover a bifurcation in infrared index (lower in MR cavities) suggesting that the molecular dichotomy is linked to residual $\mu$m-size dust within millimeter disk cavities. Put together, these results suggest a feedback process between dust depletion, gas density decrease, and molecule dissociation. Disk cavities may have a common evolutionary sequence where MR switch into MP over time. 
\end{abstract}


\section{Introduction} \label{sec: intro}
The evolution of protoplanetary disks is thought to be closely interconnected to planet formation. As protoplanets grow in the disk, when they are massive enough they dynamically open a gap around their orbital radius, a gap that can eventually separate the inner from the outer disk \citep[e.g.][]{Skrutskie1990,Marsh1992,Calvet2002,DAlessio2005}. Disk winds are also expected to open gaps and eventually lead to disk dispersal, depleting disk gas that could otherwise be accreted by protoplanets \citep[e.g.][]{alexander14,ercolano_pascucci2017,Pascucci2023}. The relative distribution of dust and gas in inner disks could in principle inform on which processes are driving disk evolution in different phases \citep[e.g.][]{espaillat14,manara23,vanderMarel23}.

Protoplanetary disks with an inner dust cavity have long attracted attention as a possibly particular phase during the global evolution from embedded objects (Class 0/I) to the final disk clearing stages (Class III and Debris Disks). This type of disk structure was first inferred from the analysis of spatially-unresolved spectral-energy-distributions (SED) of pre-main-sequence stars showing a deficit in near-infrared (NIR) continuum emission in comparison to the median SED (often taken in Taurus) followed by a steeper rise toward the far-IR. These disks were interpreted as having an optically thin inner region and an optically thick outer region, implying significant depletion of small ($\lesssim \mu$m) dust grains from a hotter inner region suggestive of an ``inside-out" disk dispersal process. These disks became first known in the literature with the terms ``transition/transitional" to highlight the evolutionary phase originally proposed as their interpretation \citep[e.g.][]{Strom1989,Skrutskie1990,koerner93,Calvet2002,DAlessio2005,Najita2007_TD,espaillat14,vanderMarel16}.

With the availability of long-baseline interferometers, inner disk dust ``cavities" or ``holes" have later been spatially resolved in millimeter continuum emission, confirming and supporting, at first, the SED-based interpretation in a handful of disks \citep[e.g.][]{Hughes07,brown09,Andrews11}, and providing new direct measurements of dust and gas clearings or depletion \citep[e.g.][]{WC_2011}. 
After the advent of ALMA, millimeter dust cavities have been spatially resolved in tens of disks, with a detection limit of $\gtrsim 5$~au in cavity size with the long-baseline array but more often lower resolution, $\gtrsim 20$~au, in large surveys of star-forming regions \citep[e.g.][]{ansdell16,pascucci2016,pinilla18,francis20}. With a larger number of spatially-resolved disk cavities, the new interferometric surveys only partially matched earlier SED detections of inner dust evolution and called for new detection, classification, and analysis procedures \citep{vanderMarel23}. Early SED-based classifications turned out to be in some cases affected by heterogeneous samples that mixed ``transitional" with  ``evolved" disks, which by showing a global SED decrease without a rise at far-IR wavelengths are now generally not considered anymore as having a dust cavity \citep{espaillat14}. Another typical source of uncertainty was the degeneracy in SED modeling due to dust properties, mass, disk geometry, and viewing angle \citep[e.g.][]{DAlessio2006,furlan09,woitke16,Ballering19,vanderMarel22}.  

On the other hand, several disks whose SED did not show any deficit in NIR continuum and therefore were not classified as ``transitional" have later been found to have a large millimeter cavity \citep{loomis17,pinilla18_CIDA,vanderMarel23}. While the NIR continuum arises from small ($\lesssim \mu$m) dust, the millimeter emission is dominated by larger ($\gtrsim$~mm) grains or particles and these two different solid populations can naturally be spatially separated by their different aerodynamic coupling to the gas \citep{zhu11,garufi13,espaillat14}. 
The wide range of cavity sizes, NIR continuum, and accretion rates indeed suggested early on that these disks might be an heterogeneous class of structures with different origins or observed in different conditions or phases of a same process \citep[e.g. planets with different masses and/or coupled to winds of different nature or dead zones, see e.g. reviews by][]{Najita2007_TD,espaillat14,ercolano_pascucci2017,vanderMarel23}.
In particular, planet-induced gaps can produce an inner disk clearing from millimeter dust grains by trapping them at larger radii while still allowing smaller dust to filtrate through the cavity, effectively producing a millimeter cavity that may or may not show a NIR deficit depending on the details of gas and dust coupling and evolution \citep[e.g.][]{rice2006,zhu2012}.

\begin{figure*}
\centering
\includegraphics[width=1\textwidth]{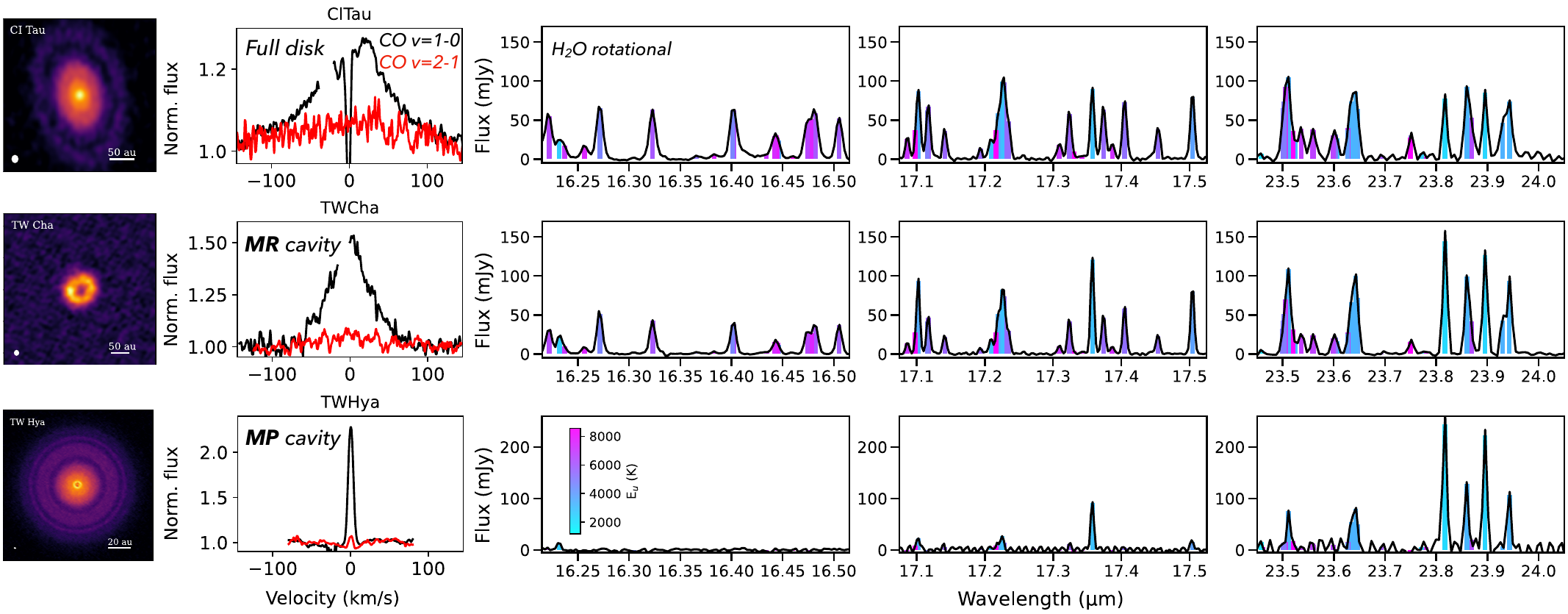} 
\caption{Inside-out depletion of CO and water vapor previously observed with ground-based and Spitzer spectra in \cite{banz17}, here reproduced with the new MIRI spectra, marking water transitions according to their upper level energy as in \cite{banz25}. Three disks are shown as examples of a full disk and the molecule-rich (MR)/molecule-poor (MP) cavity types introduced in this work. As CO lines get narrower \citep[plots to the left, with data from][]{brown13,banz22}, indicating depletion of the higher velocity hot gas at $\lesssim 0.1$~au, the higher-energy water lines (colored purple and magenta) become weaker while the lower-energy lines (blue and cyan) probing colder water at larger radii are still strong. ALMA continuum images are reported to the left for reference (see Section \ref{sec: obs}).}
\label{fig: water_seq_depl}
\end{figure*}

In terms of gas, disks with a dust cavity (identified from millimeter and/or SED studies) were found to have substantial stellar accretion rates indicating gas-rich inner regions \citep{Manara2014}, but with lower median accretion as compared to full disks in T~Tauri stars with similar dust mass \citep[][and Section \ref{sec: data}]{Najita2007_TD}. Spatially-resolved observations later demonstrated various degrees of gas depletion inside a dust cavity, generally supporting disk clearing by giant planets \citep[e.g.][]{dutrey2008,vandermarel2015,vanderMarel18}. Velocity-resolved IR and UV line profiles have shown that at least CO and \ch{H2} emit from a similar region well within the millimeter cavity \citep{rettig04,pont08,salyk09,hoadley15,doppmann17,banz22}, supporting modeling expectations for their density-dependent survival within the millimeter cavity \citep{bruderer13}. While ro-vibrational CO emission is typically found to extend within the inner dust radius from IR interferometry, suggesting CO survival within a dust-free gas-dense region, disks with a dust cavity have typically shown narrower lines implying a recession of CO to larger radii \citep{salyk09,salyk11,bp15,banz17,banz22}. The dependence of CO gas survival on residual dust within a millimeter cavity has indeed been confirmed in a large sample of disks around Herbig Ae/Be disks, by finding that CO emission recedes to larger radii with the decrease of NIR excess within the cavity \citep{banz18}. 

Spatially- and spectrally-unresolved Spitzer spectra showed that disks with a NIR deficit have lower molecular luminosity or do not show molecular detections in \ch{H2O}, HCN, and \ch{C2H2} except for CO and OH \citep{pont10,najita10,salyk11_spitz}.
Follow-up analyses found at least one large millimeter cavity with an infrared (IR) molecular spectrum very similar to a full disk \citep[DoAr~44 with a millimeter cavity of $\sim 40$~au in][]{Salyk15,salyk19}, which was attributed to residual dust within the cavity previously inferred from SED modeling \citep{espaillat10} and later spatially resolved at $\sim0.14$~au with GRAVITY \citep{Bouvier20}. A combined analysis of Spitzer and ground-based velocity-resolved spectra later showed that as CO ro-vibrational lines get narrower in disks developing an inner dust cavity, the high-energy water emission decreases while the lower-energy emission is still strong at longer wavelengths, suggesting a joint sequential depletion from hotter to colder molecular gas. We illustrate this sequential depletion in Figure \ref{fig: water_seq_depl}, which updates what previously shown with Spitzer and ground-based data in Figure 9 in \citet{banz17}. With JWST, the study of molecular gas in disk cavities is already expanding our knowledge, finding previously undetected water in PDS~70 \citep{perotti23}, a rich molecular spectrum in SY~Cha \citep{schwarz24}, and rare species like \ch{CH3+} in TW~Hya and GM~Aur \citep[][]{MINDS24,Romero-Mirza25}.

\begin{deluxetable*}{l c c c c c c c c c c c c c c}
\tabletypesize{\footnotesize}
\tablewidth{0pt}
\tablecaption{\label{tab: sample} Sample properties for cavity disks studied in this work.}
\tablehead{ Name & Dist & $T_{\rm{eff}}$ & $M_{\star}$ & $L_{\star}$ & log $L_{\rm{acc}}$ & $R_{\rm{disk}}$ & $L_{mm}$ & CO width & $R_{\rm{CO}}$  & $n_{13-26}$ & $R_{\rm{cav}}$ & Dust & Age \\ 
 & (pc) & (K) & ($M_{\odot}$) & ($L_{\odot}$) & ($L_{\odot}$) &(au) & (mJy) & (km/s) & (au) &  & (au) & cavity & (Myr) }
\tablecolumns{14}
\startdata
\multicolumn{14}{c}{Molecule rich (MR)} \\
SR~4 & 134 & 4115 & 0.61 & 1.82 & -0.35  & 31 & 65.27 & 118--55 & 0.02--0.11 & 0.75 & $< 2.5$ & IR & $0.6^{+0.1}_{-0.1}$ \\
HP~Tau & 177 & 4375 & 0.84 & 1.89 & -1.06 & 22 & 50.04 & 117--41 & 0.02--0.18  & 0.36 & $< 10$ & IR & $0.8^{+0.5}_{-0.3}$ \\
Sz~129 & 160 & 4020 & 0.73 & 0.42 & -2.15  & 76 & 99.30 & NA & NA & 0.68 & 10 & mm+IR & $3.8^{+3.3}_{-1.7}$ \\
IP~Tau & 129 & 3792 & 0.51 & 0.52 & -3.02  & 36 & 12.41 & $\sim$150--108$^{b}$ & 0.04--0.08$^{b}$ & -0.06 & 25 & mm(+IR) & $1.4^{+1.1}_{-0.6}$ \\
TW~Cha & 183 & 4020 & 0.7 & 0.50 & -1.19  & 53 & 47.05 & 132--63 & 0.05--0.23 & 0.19 & 30 & mm+IR & $2.8^{+2.3}_{-1.3}$ \\
SY~Cha & 181 & 4020 & 0.67 & 0.55 & -2.06  & 184 & 96.16 & $\sim$250--200$^{b}$ & 0.02--0.04$^{b}$ & -0.10 & 36 & mm & $2.6^{+2.2}_{-1.2}$ \\
\hline
\multicolumn{14}{c}{Molecule poor (MP)} \\
TW~Hya & 60 & 3800 & 0.61 & 0.23 & -1.78  & 58 & 105.61 & 12.9-7.3 & 0.2-0.6 & 1.19 & 2.4 & mm+IR & $5.6^{+5.6}_{-2.7}$\\
HD~143006 & 167 & 4870 & 1.48 & 2.61 & -1.10  & 81 & 84.29 & 28--22 & 1.68--2.55 & 1.55 & 6 & mm+IR & $1.7^{+1.2}_{-0.7}$\\
T~Cha & 103 & 5300 & 1.76 & 3.2 & -2.73  & 63 & 56.83 & \nodata$^{a}$ & \nodata$^{a}$ & 2.35 & 36 & mm+IR & $2.9^{+2.2}_{-1.2}$\\
GM~Aur & 141 & 4115 & 0.69 & 0.91 & -0.38  & 160 & 175.68 & 77--42$^{b}$ & 0.26--0.89$^{b}$ & 2.29 & 40 & mm+IR & $1.3^{+0.8}_{-0.5}$\\
RY~Lup & 158 & 4710 & 1.27 & 1.84 & -0.92 & 135 & 109.67 & 99--25 & 0.29--4.6 & 0.65 & 67 & mm+IR & $1.9^{+1.4}_{-0.8}$\\
PDS~70 & 112 & 4138 & 0.76 & 0.35 & $<$~-3.07  & 110 & 37.12 & NA & NA & -0.22 & 74 & mm & $4.5^{+3.8}_{-2.0}$ \\
\enddata
\tablecomments{
Targets are separated into MR and MP following this work, and are ordered by millimeter cavity size $R_{\rm{cav}}$ in each group. References: Distances are from Gaia DR3 parallaxes \citep{gaia_mission,gaiaDR3}, stellar properties are from \citet{manara23}, which reports $T_{\rm{eff}}$ from \citet{hh14}, except for those not included in that work: TW~Hya from \citet{fang18}, PDS~70 from \citet{skinner2022}, and T~Cha from \citet{alcala97}. $L_{\rm{acc}}$ is derived in this work from HI lines observed with MIRI using relations from \citet{Tofflemire25}. The infrared index $n_{13-26}$ is measured from the MIRI spectra in this work (Appendix \ref{app: IRindex}). $R_{\rm{disk}}$ (the disk radius including 90--95\% of the millimeter emission depending on what reported in different works) and disk inclinations are from \citet{huang18,long19}. The luminosity at 1.3~mm, $L_{mm}$, is obtained by scaling the mm flux at 140~pc as done in \citet{andrews18}, with fluxes from \citet{ansdell16,lommen07,dsharp,macias21,long19,Orihara2023,fasano25} and Long et al. in prep. Ground-based near-infrared CO line widths observed with iSHELL and CRIRES \citep{brown13,banz22} are measured at the 10\% and 50\% of the line peak from the stacked profiles (the values given in the table are the half line width velocities at 10\% and 50\%, or FW10\% - FW50\%); $R_{\rm{CO}}$ is the Keplerian radius from these velocities. $^{a}$: in the case of T~Cha, CO lines detected in CRIRES spectra include both emission and absorption and do not provide an estimate of the emission line profile \citep{brown13}. $^{b}$: in these disks, CRIRES and NIRSPEC observe a stellar photospheric spectrum with weak and broad emission in SY~Cha and narrow emission in GM~Aur \citep[see spexodisks.com,][]{spexodisks}; the line widths in GM~Aur are adopted from \cite{salyk09} and in IP~Tau from \cite{doppmann17}. $R_{\rm{cav}}$ is the cavity radius as determined from spatially-resolved ALMA images taken at the peak of the first dust ring limiting the cavity, as reported in \cite{huang18,huang20}, \cite{francis20}, \cite{vanderMarel23} and references therein, except for TW~Hya from \cite{andrews16} and TW~Cha from F. Long (priv. comm.). The stellar age was determined using the tool from \cite{Deng2025} based on the IDL code developed by \cite{pascucci2016}, which uses evolutionary tracks of \citet{Feiden16} for sources with $T_{\rm eff}>3900\mathrm{~K}$ and that of \citet{Baraffe15} for $T_{\rm eff} \le 3900\mathrm{~K}$.
}
\end{deluxetable*}

\begin{figure*}
\centering
\includegraphics[width=1\textwidth]{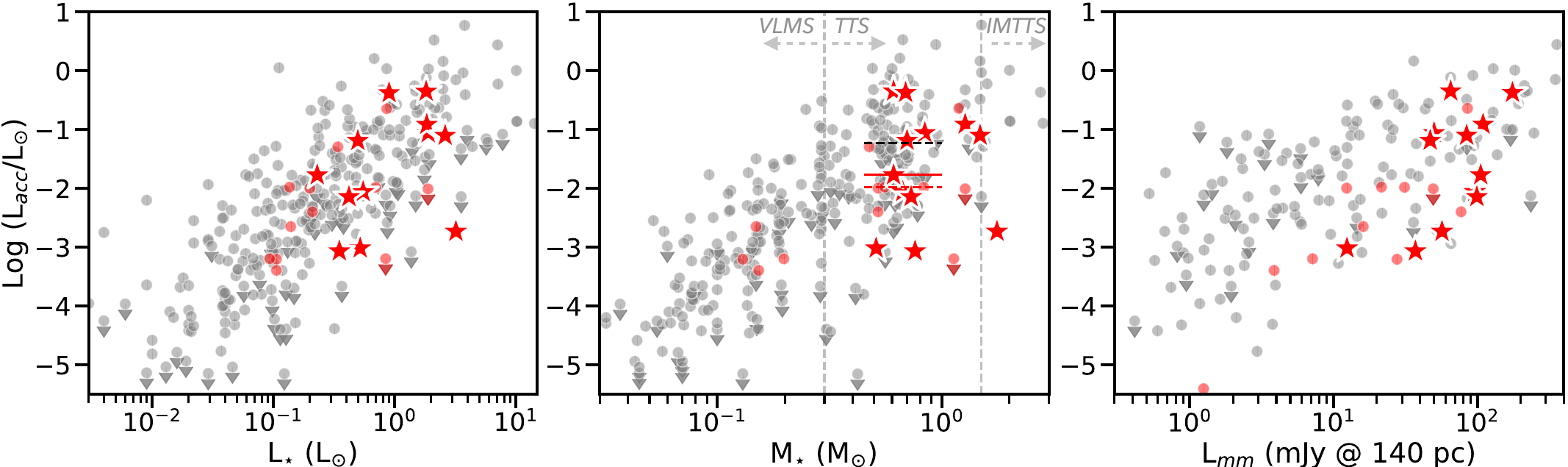} 
\caption{Comparison between the sample of cavity disks included in this work (red stars) and the large protoplanetary disk sample compiled in \citet{manara23} in grey and red dots. $L_{\rm{acc}}$ values for this work are estimated from MIRI spectra (Section \ref{sec: analysis_1}). Disks labeled in \citet{manara23} as having a dust cavity are marked in red. The mm luminosity L$_{mm}$, taken as the flux density normalized to 140~pc as in \citet{andrews18}, is taken at 1.3~mm, where the emission is more optically thin than at 0.89~mm and should better reflect the dust mass. The approximate boundaries between T~Tauri stars (TTS), very-low-mass stars (VLMS), and intermediate-mass T~Tauri stars (IMTTS) are marked for reference \citep{luhman10,calvet04}. The horizontal lines at stellar masses of 0.45--1~$M_{\odot}$ show the median log $L_{\rm{acc}}$ value for full disks (black), the cavity disks from \citet{manara23} (dashed red) and the cavity disks in this work (solid red).}
\label{fig: Manara_compar}
\end{figure*}

In this work, we assemble a first sample of disks with a dust cavity (see Section \ref{sec: sample} for details on the selection) that have been observed with JWST-MIRI and expand on previous results to investigate molecular survival specifically in the disks of young pre-main-sequence stars of solar mass (T~Tauri stars). We find that the high-quality MIRI spectra clearly show a dichotomy in molecular emission in these disks, which we use to formally define two types of inner disk cavities (Section \ref{sec: analysis}): molecule-rich (MR) and molecule-poor (MP). Bearing in mind that the sample in this work is still small (12 objects) and thus requires further investigations with larger samples \citep[see comparison to][in Figure \ref{fig: Manara_compar}]{manara23}, we start in this work to identify the specific properties of MR and MP cavities and discuss them in the context of different levels of dust filtration through the millimeter cavity and of a gas density decrease, which we discuss in the context of a common evolution or different origins (Section \ref{sec: disc}).

In this work, we refrain from using the original interpretation-based term ``transitional" and we use instead the observation-based terms of ``mm-cavity" disk if it has a spatially-resolved millimeter dust cavity and ``IR-cavity" disk if it has the infrared index $n_{13-26} > 0$ that may indicate depletion of smaller dust in the inner disk (see Section \ref{sec: data} and Appendix \ref{app: IRindex}). We use ``cavity disk" as a general term and ``mm+IR cavity" disk to indicate that it has a cavity detected in both tracers. Protoplanetary disks without detected deficits indicative of an inner dust cavity in either tracer are identified as ``full" disks and are used as reference to analyze what observed in disks with a dust cavity. These full disks may have gaps at multiple radii that are observationally different from a dust cavity in being (generally) radially narrower and having a large spatially-resolved optically thick inner disk in millimeter emission. These disk structures might still be sequential in terms of evolution, as it has been proposed that planet-induced gaps may later develop into cavities in the context of the formation of giant planets through core accretion \citep{dodson11,cieza2021,Orcajo2025}.

\section{Sample \& Observations} \label{sec: data}
\subsection{Sample definition and properties} \label{sec: sample}
The selection of 12 disks for this work is based on Class II protoplanetary disks that have an identified dust cavity and are available from a number of Cycle 1 and 2 JWST-MIRI programs (see Section \ref{sec: obs}). Sample properties, including the cavity type in each disk, are reported in Table \ref{tab: sample} and are visualized in Figure \ref{fig: Manara_compar} to put the sample into a broader context, illustrating that it is composed of stars with mass of 0.5--1.5~$M_{\odot}$ with relatively mm-bright disks. The cavity identification is based on ALMA images complemented with the slope at infrared wavelengths as measured directly from the JWST-MIRI continuum, both of which have their own limitations in revealing a disk cavity (see Section \ref{sec: intro} and Appendix \ref{app: IRindex}). The selection of mm-cavity disks is based on high-resolution ALMA images, which directly resolve cavities in millimeter dust continuum emission; the limitation of this selection is that ALMA images can still miss small cavities with size $< 5$--20~au depending on the spatial resolution achieved in nearby star-forming regions at 120--200~pc in different works (Section \ref{sec: intro}). In this work, 10/12 disks have a spatially-resolved disk cavity in millimeter emission (Table \ref{tab: sample}).

The second type of data we use is the JWST-measured infrared index $n_{13-26}$ (formerly the $n_{13-30}$ index when measured from Spitzer spectra) that measures the slope of the SED between 13 and 26~$\mu$m and is sensitive to the depletion of small ($\lesssim \mu$m) dust in the inner hot disk region \citep{brown07,furlan09}. While different limits have been adopted in the past to identify a dust cavity based on the IR index, based on spatially-resolved cavities and SED modeling \cite{banz20} discussed that an inner dust cavity may be revealed by $n_{13-30} \gtrsim 0$, now $n_{13-26} \gtrsim 0$ with MIRI (see Appendix \ref{app: IRindex}). In this work, 9/12 disks have an IR cavity based on their positive $n_{13-26}$ index (plus two border-line cases with negative index of $\sim -0.1$), of which 7/12 have both a mm and a IR deficit (mm+IR cavity, Table \ref{tab: sample}). It should be considered that the infrared SED (and therefore the IR index) is known to be variable in disks that have a cavity \citep{espaillat11,espaillat24,Xie_2025}, which will produce scatter in indices measured at different times (see Appendix \ref{app: IRindex}). An important example to mention is the case of PDS~70 (discussed in Appendix \ref{app: additional}), which in this work shows only a mm cavity by having a negative IR index, but using a previous Spitzer spectrum it would have a positive index and be a mm+IR cavity \citep{perotti23,Jang2024}.

Disk models show that the SED at 10--30~$\mu$m is sensitive to the removal of $\lesssim \mu$m dust in a region as small as $\sim 1$--2~au \citep[e.g.][]{woitke16,Ballering19} and therefore can reveal small cavities that ALMA may not be able to resolve. We include in this sample two disks that might be in this situation: SR~4 and HP~Tau, which have significantly positive $n_{13-26} =$ 0.36--0.75 but no cavity detected with ALMA at the current spatial resolution, providing an upper limit of $\lesssim 2.5$~au and $\lesssim 10$~au for a dust cavity respectively \citep[taking the upper limit to be given by FWHM/2 of the synthesized beam obtained in each case,][]{huang18,long19}. For comparison, TW~Hya has a 2.4~au cavity, which is spatially resolved due to its much closer distance of only 60~pc \citep{andrews16}, and has an index $n_{13-26} \sim 1.2$ ($\sim 2$--3 times larger than in HP~Tau and SR~4). In case they do not have a dust cavity, HP~Tau and SR~4 must have another explanation for their positive IR index that mimics a cavity. A high disc inclination can mimic a cavity \citep[e.g.][]{Ballering19,vanderMarel22} but this possibility can be excluded in this case since these disks are almost face-on \citep{dsharp,long19}. An alternative explanation proposed early on for positive IR indices in the range $\sim$~0--0.5 is a dust/gas mass ratio closer to standard ISM values in the disk surface (dust/gas = 0.01--0.0001), while most Class II disks should be dust-depleted by settling \citep[dust/gas = 0.0001--0.00001][]{DAlessio2006,furlan06,furlan09}. If these two disks are less settled or have dust/gass closer to ISM values, it may provide a reason for their positive index.
In conclusion, in this work we still include HP~Tau and SR~4 in the sample of cavity disks by assuming they do have an unresolved dust cavity and we adopt an upper limit on their cavity size as written above.

\subsection{Observations} \label{sec: obs}
All disks were observed over the full wavelength coverage of 4.9--28\,$\mu$m with the Medium Resolution Spectrometer \citep[MRS,][]{jwst-mrs} mode on MIRI \citep[][]{miri,miri2}. The data come from the following programs, mostly from the JDISCS \citep{pontoppidan24,Arulanantham25} and MINDS \citep{minds_kamp,MINDS24} collaborations: 3 disks from GO-1584 (PI: C. Salyk; co-PI: K. Pontoppidan), 3 disks from GO-1282 (PI: T. Henning; co-PI: I. Kamp), 2 disks from GO-1640 (PI: A. Banzatti), 1 disk from GO-1549 (PI: K. Pontoppidan), 1 disk from GO-2025 (PI: Karin \"Oberg), 1 disk from GO-2260 (PI: I. Pascucci), 1 disk from GO-3228 (PI: I. Cleeves) for a total of 12 disks. All these MIRI spectra were previously published first in the following papers: \citet{perotti23,banz23b,schwarz24,pontoppidan24,munozromero24a,MINDS24,bajaj24,Sellek24,banz25,Arulanantham25,Romero-Mirza25}, except for IP~Tau from Romero-Mirza et al. in prep.

All MIRI-MRS spectra were extracted and wavelength-calibrated with the JDISCS pipeline as described in \cite{pontoppidan24}, which adopts the standard MRS pipeline \citep{MIRI_pip} up to stage 2b producing the three-dimensional cubes and then uses observed asteroid spectra from GO program 1549 as calibrators to provide high-quality fringe removal and characterization of the spectral response function to maximize S/N in channels 2--4. A standard star is used in channel 1 where asteroid spectra have low S/N. To ensure similar fringes and maximize the quality of their removal, target acquisition with the MIRI imager was adopted to reach sub-spaxel precision in placing science target and asteroids on the same spot on the detector. For the data included in this work, we used MRS pipeline version 12.1.5 with Calibration Reference Data System context jwst\_1364.pmap and JDISCS reduction version 9.0 that includes an improved bad pixel correction and flux calibration uncertainty down to 1--2\% across MIRI.
Before the analysis of molecular lines, the MIRI spectra were continuum-subtracted with the procedure presented in \cite{pontoppidan24} updated to apply a wavelength-dependent offset correction based on line-free regions to account for gas absorption where present as described in \cite{banz25}. The MIRI spectra are illustrated in Appendix \ref{app: sample_figures}.

In addition to MIRI spectra, we include dust measurements of inner dust cavities as measured in millimeter continuum with ALMA (see references in Table \ref{tab: sample}). To illustrate the dust emission radial structures we include some ALMA continuum images with data from \citet{andrews16,dsharp,long19,huang20,benisty21}, and Long et al. in prep., e.g. in Figure \ref{fig: water_seq_depl}. To complement the limited kinematic information provided by MIRI spectra, we also include velocity-resolved M-band rovibrational CO emission observed with ground-based infrared spectrographs \citep[iSHELL and CRIRES,][]{ishell22,crires} from \citet{brown13,banz22}. Details on the data reduction and analysis of these datasets can be found in the original papers.

\begin{figure*}
\centering
\includegraphics[width=1\textwidth]{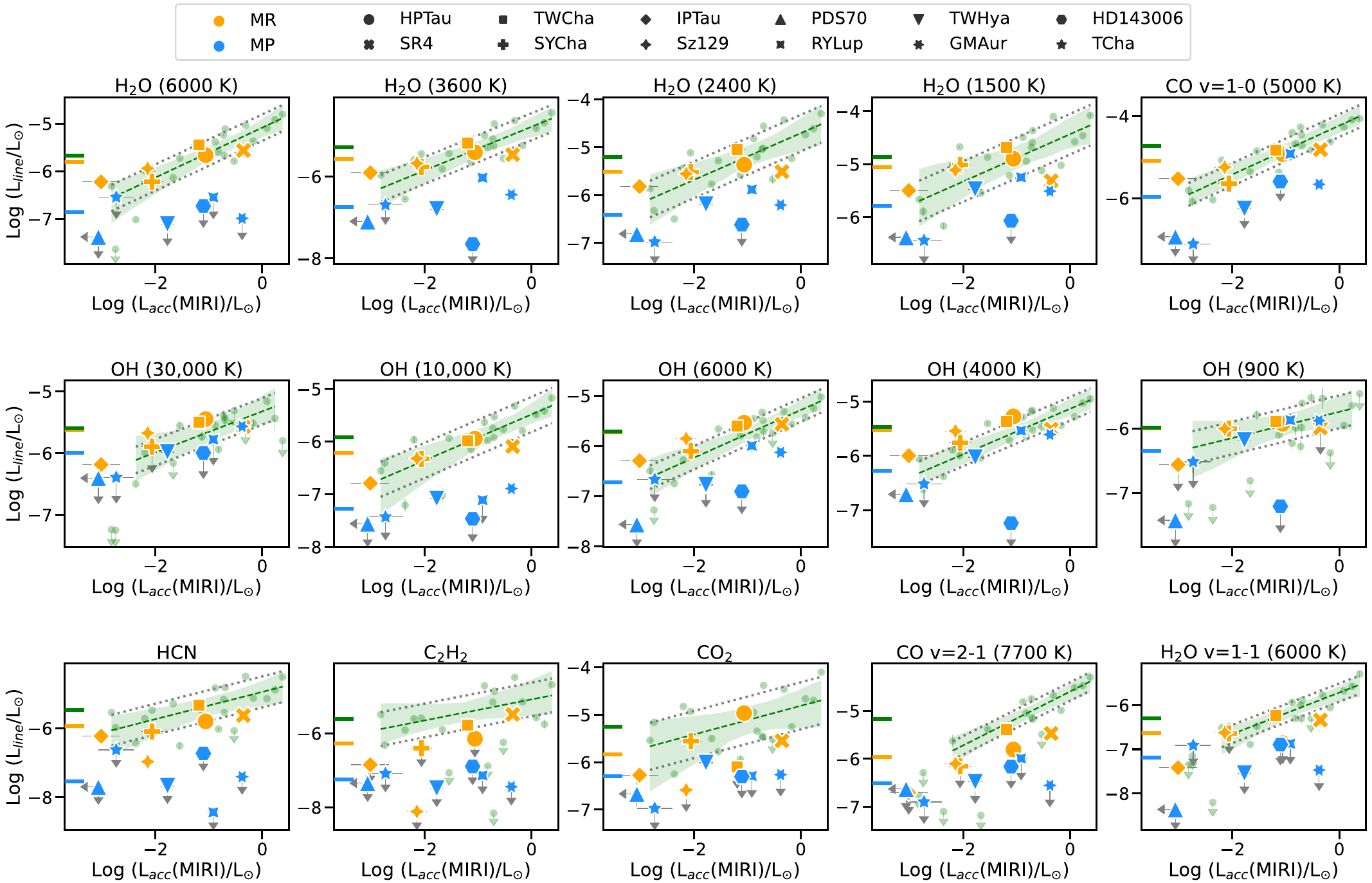} 
\caption{Line luminosity of multiple molecules as a function of accretion luminosity as measured in MIRI spectra. Where included, the number in Kelvin indicates the approximate upper level energy (see Table \ref{tab: lin_fit_params}). The sample of disks with cavities is separated into ``molecule-rich" (MR, orange) and ``molecule-poor" (MP, blue) as defined in this work (see Section \ref{sec: analysis_1}). The reference sample shown in green is composed of full disks in the JDISCS-C1 sample (Section \ref{sec: analysis_1}), with linear fits and their 95\% confidence intervals shown as dashed lines and shaded regions and regression parameters included in Appendix \ref{app: new_lines_4_corr}. The dotted lines show one standard deviation of the distribution of full disks around each correlation, which is used to define MR cavities in the higher-energy water lines (Section \ref{sec: analysis}). Median luminosity values for full, MR, and MP disks are shown as bars to the left in each panel. Higher-vibrational lines of CO and \ch{H2O} are included in the bottom row of plots for reference to the sub-luminous organics (see discussion in Section \ref{sec: gas density}).}
\label{fig: lum_correl}
\end{figure*}

\section{Analysis \& Results} \label{sec: analysis}

\subsection{Defining molecule-rich and molecule-poor cavities} \label{sec: analysis_1}
Building on previous work that found a few exceptions to a general trend of low molecular detections in disks with dust cavities (see Section \ref{sec: intro}), in this work we use the higher-resolution MIRI spectra and new accretion estimates from MIRI to clearly define a classification of disk cavities based on their molecular spectra. 
As reference for the observed gas properties in inner disks, we use the T~Tauri disk sample from the JDISC Survey obtained in Cycle 1 \citep{pontoppidan24,Arulanantham25}, after removing disks with cavities that are studied in this work (TW~Cha, RY~Lup, Sz~129, SR~4, HP~Tau). We label this sample as ``JDISCS-C1" in this work. This reference sample of molecule-rich full disks defines tight global trends and correlations with the accretion luminosity (Figure \ref{fig: lum_correl}), which previous work found to be the strongest parameter driving the observed molecular line luminosities in T~Tauri stars \citep[][]{banz20,banz25}.

In Figure \ref{fig: lum_correl}, we present the sample of cavity disks as compared to the correlations defined by full disks with the accretion luminosity $L_{\rm{acc}}$, which we measure from MIRI spectra using the three HI lines (10-7, 7-6, 8-7) following \cite{Tofflemire25}. The value we adopt is the weighted average of $L_{\rm{acc}}$ from these three lines where detected and we show it as $L_{\rm{acc}}$(MIRI) in the figures in this work. While Balmer-jump fits typically provide the most reliable $L_{\rm{acc}}$ estimates, here we use $L_{\rm{acc}}$ estimates from MIRI HI lines because they provide an accretion tracer simultaneous to the MIRI gas lines and because they give us $L_{\rm{acc}}$ for the entire sample used in this work. For reference, we note that if we use literature values for $L_{\rm{acc}}$, available from \cite{manara23} only for part of the sample and complemented with other papers where necessary, the trends and cavity disk dichotomy reported in Figure \ref{fig: lum_correl} do not change despite a slightly larger scatter that we attribute to time variability of $L_{\rm{acc}}$ \citep[see also discussion of this topic in e.g.][]{claes22,smith_2025}. In this analysis, we prefer to use the simultaneous HI-derived $L_{\rm{acc}}$ even if they may have more uncertain luminosity relations \citep[][]{Tofflemire25,Fiorellino2025,Shridharan2025}, or be potentially affected in some disks by jet contamination \citep{bajaj24}.

\begin{deluxetable*}{l c c c c c| c c c c c | c c c c c}
\tablecaption{\label{tab:detections} Detection summary for molecular tracers in cavity disks analyzed in this work.}
\tablehead{ Name & \multicolumn{5}{c}{H$_2$O} & \multicolumn{5}{c}{OH} & HCN & C$_2$H$_2$ & CO$_2$ & CO & CO \\
& 6000 & 3600 & 2400 & 1500 & $v$=1-1 & 30,000 &
10,000 & 6000 & 4000 & 900 &  &  &  & $v$=1-0 & $v$=2-1
}
\startdata
\multicolumn{16}{c}{Molecule rich (MR)} \\
\hline
SR~4    & $\surd$ & $\surd$ & $\surd$ & $\surd$ & $\surd$ & $\surd$ & $\surd$ & $\surd$ & $\surd$ & -- & $\surd$ & $\surd$ & $\surd$ & $\surd$ & $\surd$ \\
HP~Tau  & $\surd$ & $\surd$ & $\surd$ & $\surd$ & -- & $\surd$ & $\surd$ & $\surd$ & $\surd$ & $\surd$ & $\surd$ & -- & $\surd$ & $\surd$ & $\surd$ \\
Sz~129  & $\surd$ & $\surd$ & $\surd$ & $\surd$ & $\surd$ & $\surd$ & $\surd$ & $\surd$ & $\surd$ & $\surd$ & $\surd$ & -- & -- & $\surd$ & $\surd$ \\
IP~Tau  & $\surd$ & $\surd$ & $\surd$ & $\surd$ & $\surd$ & $\surd$ & $\surd$ & $\surd$ & $\surd$ & -- & $\surd$ & -- & $\surd$ & $\surd$ & -- \\
TW~Cha  & $\surd$ & $\surd$ & $\surd$ & $\surd$ & $\surd$ & $\surd$ & $\surd$ & $\surd$ & $\surd$ & $\surd$ & $\surd$ & $\surd$ & -- & $\surd$ & $\surd$ \\
SY~Cha  & $\surd$ & $\surd$ & $\surd$ & $\surd$ & -- & -- & $\surd$ & $\surd$ & $\surd$ & $\surd$ & $\surd$ & -- & $\surd$ & $\surd$ & -- \\
\hline
\multicolumn{16}{c}{Molecule poor (MP)} \\
\hline
TW~Hya   & -- & $\surd$ & $\surd$ & $\surd$ & -- & $\surd$ & $\surd$ & -- & $\surd$ & $\surd$ & -- & -- & $\surd$ & ($\surd$) & --  \\
HD~143006 & -- & -- & -- & -- & -- & -- & -- & -- & -- & -- & -- & -- & -- & ($\surd$) & --  \\
T~Cha    & -- & -- & -- & -- & -- & -- & -- & -- & -- & -- & -- & -- & -- & ($\surd$) & --  \\
GM~Aur   & -- & $\surd$ & $\surd$ & $\surd$ & -- & $\surd$ & $\surd$ & $\surd$ & $\surd$ & $\surd$ & -- & -- & -- & $\surd$ & --  \\
RY~Lup   & -- & $\surd$ & $\surd$ & $\surd$ & -- & -- & -- & $\surd$ & $\surd$ & $\surd$ & -- & -- & -- & $\surd$ & --  \\
PDS~70   & -- & $\surd$ & $\surd$ & $\surd$ & -- & -- & -- & -- & $\surd$ & -- & ($\surd$) & ($\surd$) & $\surd$ & -- & --  \\
\enddata
\tablecomments{Targets are ordered from smaller to larger mm cavity size as in Table \ref{tab: sample}. Lines are considered detected if the measured line flux is greater than 2$\sigma$. See Appendix \ref{app: new_lines_4_corr} for line definitions, spectral ranges, and line flux and error measurements. In the case of \ch{H2O} and \ch{OH} lines, the column header reports the upper level energy in K as shown in Figure \ref{fig: lum_correl}. CO is detected in TW~Hya, HD~143006, and T~Cha with ground-based spectrographs, but not with MIRI. See Appendix \ref{app: additional} for tentative detections in PDS~70.}
\end{deluxetable*}

First, we report here the results for the reference JDISCS-C1 sample (green points and linear regressions in Figure \ref{fig: lum_correl}) of full disks for a number of molecular lines. We adopt the line list defined in \cite{banz25} for \ch{H2O} and CO and we add specific lines and ranges to measure OH across energy levels (similarly to what was done for the water lines) and organic emission as explained in Appendix \ref{app: new_lines_4_corr}. The line luminosity of all of these molecules correlates significantly with log$L_{\rm{acc}}$ and less with log$L_{\rm{\star}}$ (see regression results in Table \ref{tab: lin_fit_params}), confirming previous results from Spitzer \citep{banz17,banz20}, and shows stronger correlation and steeper slopes for higher upper level energy, as recently found for \ch{H2O} lines from MIRI spectra \citep{banz23b,banz25}. The organics show larger scatter but similar trends, with stronger correlation (excluding upper limits) and steeper slope for HCN and \ch{C2H2} in comparison to \ch{CO2}\footnote{In the case of \ch{CO2}, part of the scatter could be due to residual contamination, see Appendix \ref{app: new_lines_4_corr}}, in line with their relative excitation temperatures that are found to be generally lower for \ch{CO2}, suggesting that it traces cooler gas \citep{Arulanantham25}. Thermo-chemical models explain these correlations as a dependence of \ch{H2O} and OH emission on UV luminosity, which dominates $L_{\rm{acc}}$, while organic emission should mostly depend on X-ray luminosity instead \citep[][see also Section \ref{sec: gas density}]{Woitke24}, consistent with the lower correlation and larger scatter in their trends with $L_{\rm{acc}}$.

Once these global trends are defined by the reference sample of full disks, comparison to disks with cavities in Figure \ref{fig: lum_correl} demonstrates the dichotomy that is the main focus of this work. We base the classification on the \ch{H2O} higher-energy lines (3600--6000~K, top row of plots in the figure), which are some of the most reliable, high S/N lines and probe the inner hotter molecular gas on a region that has been found in previous work to reflect the formation of a inner disk cavity \citep[][and Figure \ref{fig: water_seq_depl}]{banz17,salyk19}. Specifically, we measure the deviation of cavity disks from the best-fit correlation and compare it to the standard deviation of the distribution of full disks around the best-fit line. Half of the disk cavity sample in this work is consistent with full disks within 1 standard deviation in both the 3600~K and 6000~K \ch{H2O} lines, making them undistinguishable from a full disk; we call these ``molecule-rich" (MR) cavities (see e.g. the example of TW~Cha in Figure \ref{fig: water_seq_depl}) and mark them in orange in all figures. 

The other half of the sample is consistently sub-luminous in most other molecular lines, especially in the hot (T $> 400$~K) molecular gas tracers. We call these ``molecule-poor" (MP) cavities  (see e.g. example of TW~Hya in Figure \ref{fig: water_seq_depl}) and mark them as light blue datapoints in all figures. In other words, the distinction we define here with MR and MP is between a disk cavity that has molecular emission as luminous as a full disk that has similar stellar and accretion luminosity, or rather it is significantly sub-luminous (see Table \ref{tab:detections} for a detection summary) indicating molecular depletion as we will discuss in Section \ref{sec: disc}. This dichotomy is visible but reduced in the colder gas tracers, especially the lower-energy \ch{H2O} and OH lines. This difference in the depletion of higher versus lower-energy water lines confirms what was previously observed from the combination of Spitzer and CRIRES data, where water emission in cavity disks was observed to reduce from higher to lower-energy levels suggesting depletion from hotter to colder molecular gas in an inside-out fashion \citep[Figure 9 in][and Figure \ref{fig: water_seq_depl} in this work]{banz17}. 

\begin{figure*}
\centering
\includegraphics[width=1\textwidth]{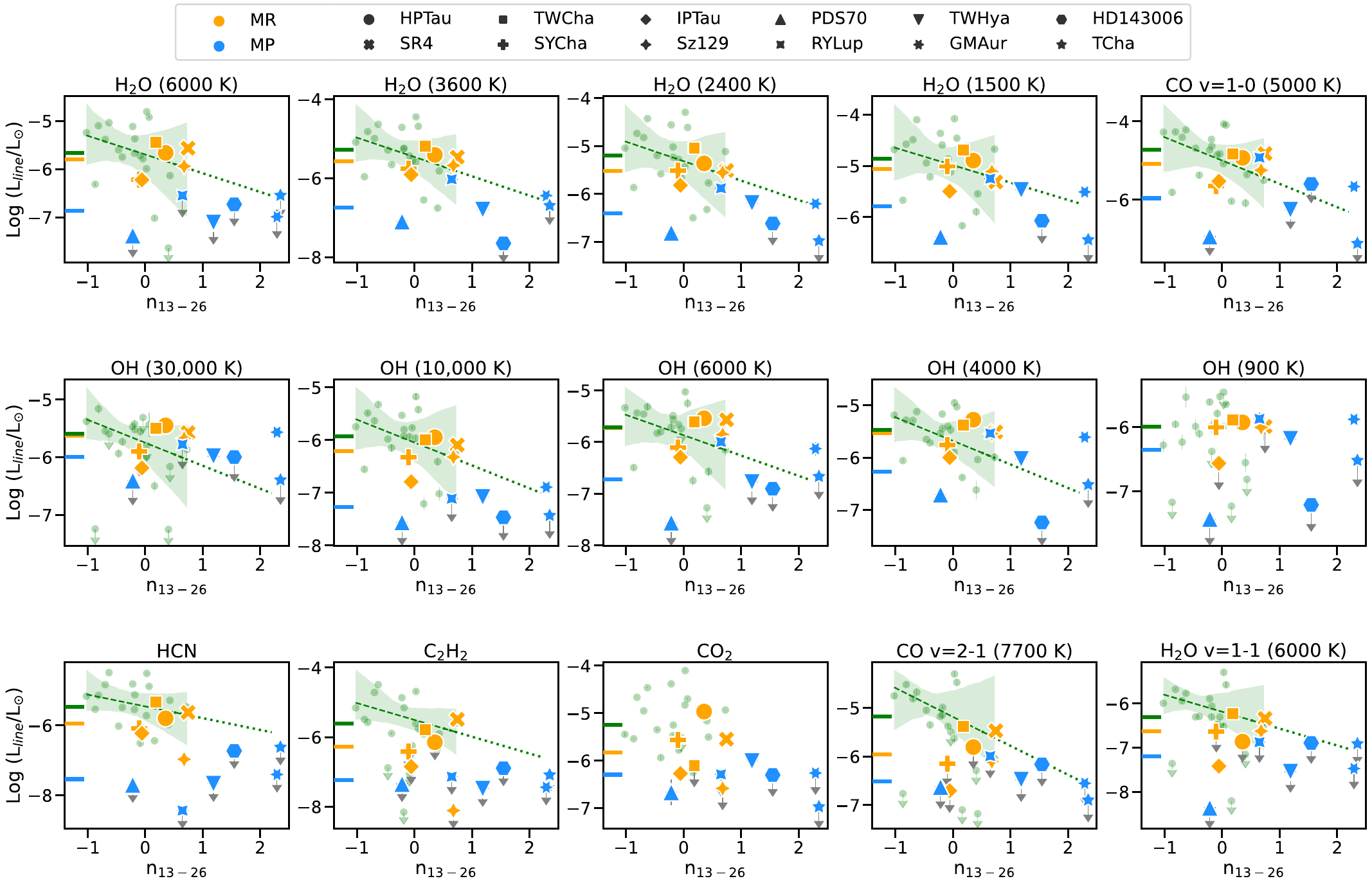} 
\caption{Same as Figure \ref{fig: lum_correl} but showing the infrared index $n_{13-26}$, the slope in the SED between 13 and 26~$\mu$m. Linear fits to the full disks and their 95\% confidence intervals are shown as dashed lines and shaded regions when considered detected (absolute value of the Pearson coefficient $> 0.25$), with regression parameters included in Table \ref{tab: lin_fit_params}. Dotted lines show extrapolations of full-disks linear fits when a correlation is detected in the full disks, for comparison to the distribution of cavity disks.}
\label{fig: lum_correl_IRindex}
\end{figure*}

It's worth noting some important exceptions.
Organic emission is on average sub-luminous in all cavities including the MR, though not as extremely as in the MP cavities (Figure \ref{fig: lum_correl}). This is especially visible in \ch{C2H2}, which is firmly detected only in two MR cavities. A drop in line luminosity is also observed in the CO $v$=2-1 and, to a lesser extent, the \ch{H2O} $v$=1-1 lines, which tend to be on average sub-luminous in all disk cavities, including MR. This will be discussed in Section \ref{sec: gas density}.
The OH lines, instead, show distinct behavior as a function of upper level energy across the MIRI spectrum. The highest-energy lines populated by prompt emission following water photodissociation ($E_u > 20,000$~K emitting at $< 11$~$\mu$m) are slightly super-luminous in MR cavities, and consistent with full disks in MP (but only detected in GM~Aur and TW~Hya). OH lines with $E_u$ of 6000--10,000~K are consistent with full disks in MR cavities, and sub-luminous in MP cavities. Going to lower energies, OH lines with $E_u = 4000$~K are again slightly super-luminous in MR and slightly sub-luminous in MP cavities as compared to full disks. The lowest-energy OH lines covered with MIRI, the $E_u \sim 900$~K lines at 24.64~$\mu$m, show that MR cavities have luminosity generally consistent with the full disks.

\begin{figure*}
\centering
\includegraphics[width=1\textwidth]{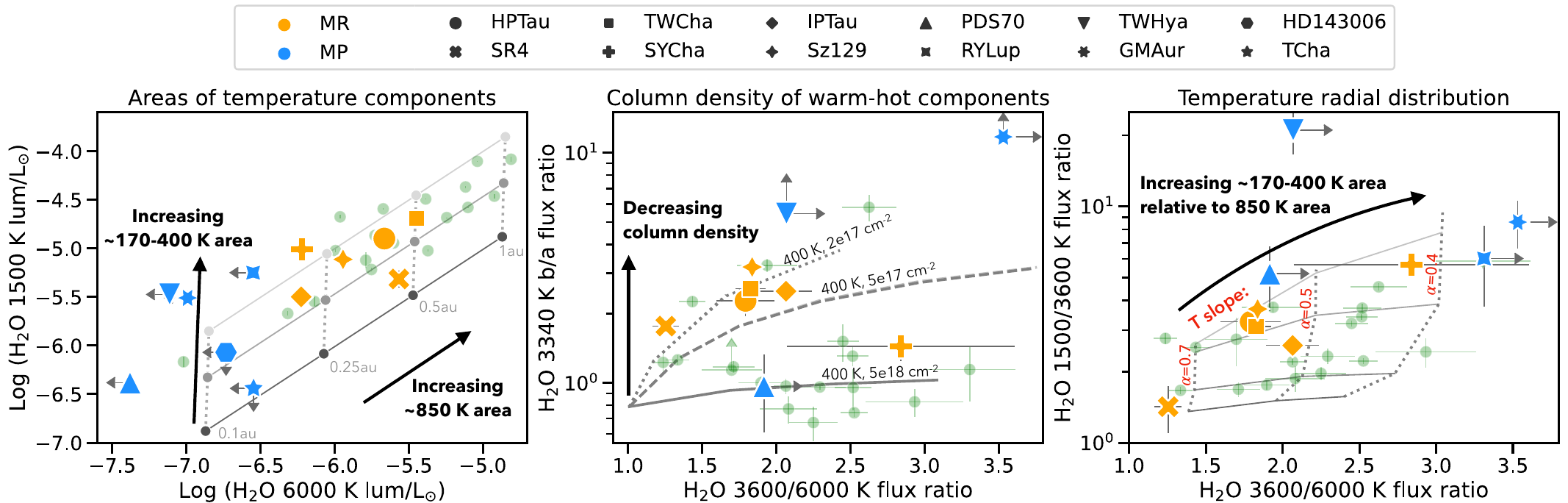} 
\caption{Water diagnostic diagrams defined in \cite{banz25}, with arrows to indicate the main directions for simple empirical interpretations based on multi-component slab models in LTE. As in Figure \ref{fig: lum_correl}, MR cavities have 6000 K \ch{H2O} luminosities comparable with those measured in full disks, while MP cavities are significantly sub-luminous (left plot). Cavity disks in general show a lower column density of the warm-hot water components (middle plot). The diagnostic line ratios in MR cavities show a similar temperature slope as in full disks but show a relatively stronger colder water reservoir (right plot). MP cavities tend to cluster at the top right of the plot due to their decreased hot water luminosity.}
\label{fig: water_diagrams}
\end{figure*}

\begin{figure}
\centering
\includegraphics[width=0.48\textwidth]{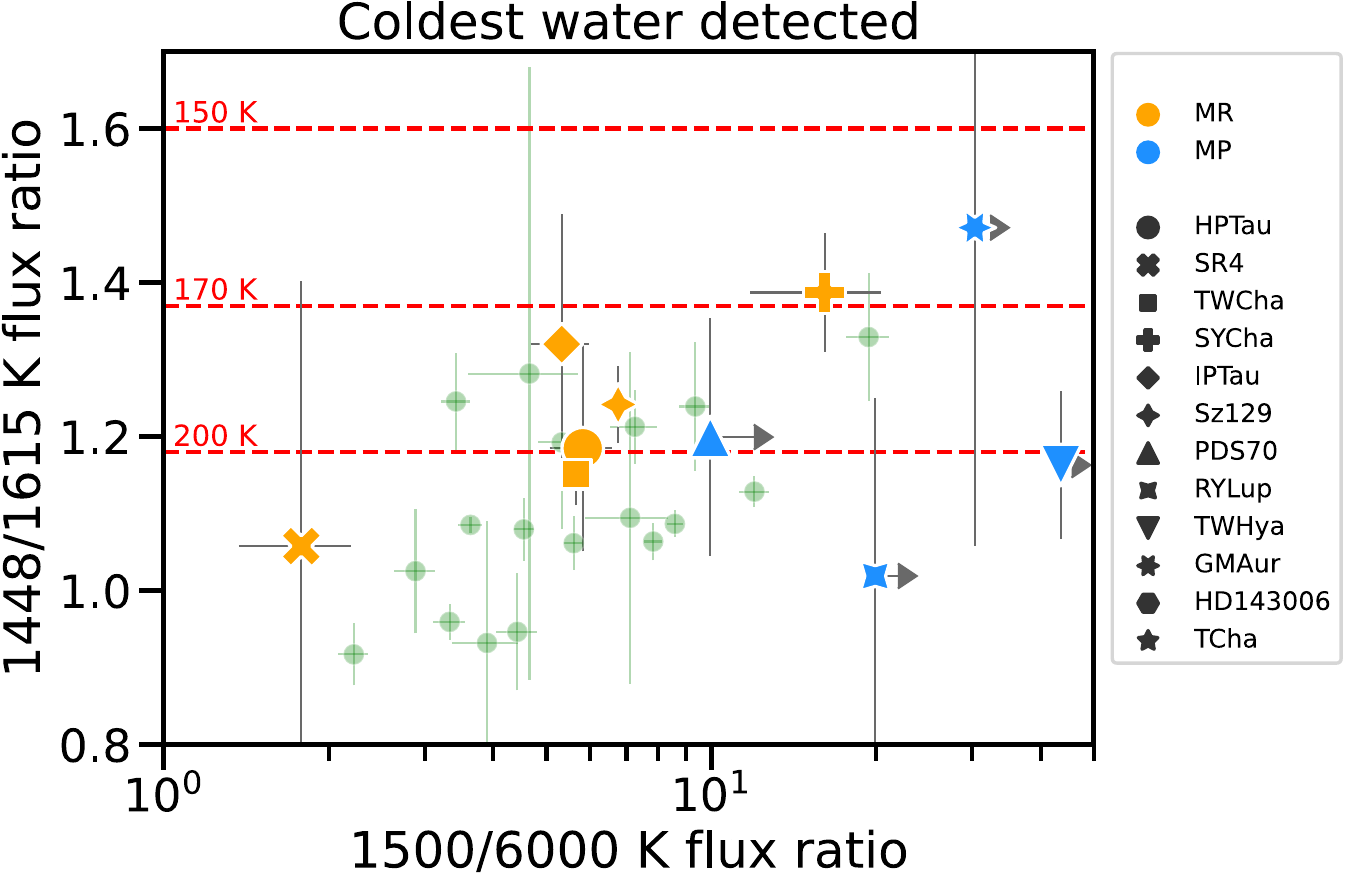} 
\caption{Cold water diagnostic diagram defined in \cite{banz25} based on the line flux ratio between the two low-energy lines near 23.85~$\mu$m. Excitation temperatures labeled in red assume LTE and a column density of $10^{17}$~cm$^{-2}$. Cavity disks tend to have larger line ratios on both axes as indicative of colder water emission than in full disks, with MP cavities clustering at the right of the plot due to the decreased hot water luminosity as in Figure \ref{fig: water_diagrams}. }
\label{fig: cold_water_diagr}
\end{figure}

Figure \ref{fig: lum_correl_IRindex} follows Figure \ref{fig: lum_correl} but shows molecular luminosities as a function of infrared index $n_{13-26}$ as measured in MIRI spectra (Appendix \ref{app: IRindex}). The trends between molecular luminosity and $n_{13-26}$ generally confirm results from Spitzer previously reported in \cite{salyk11_spitz,banz20}: molecular luminosities have an anti-correlation with $n_{13-26}$ that is consistent with a linked depletion of inner disk small dust grains and molecular gas. In this work, we improve previous Spitzer results in two ways. First, full disks on their own show in many cases (especially in higher-energy lines, see results in Table \ref{tab: lin_fit_params}) anti-correlations between the line luminosity and $n_{13-26}$ even extending to negative values of the index (which may correspond to different levels of dust settling in full disks, see Appendix \ref{app: IRindex} and Section \ref{sec: disc}). Exceptions are the 900~K OH lines and \ch{CO2}, where a correlation is not detected in the full disks alone. Second, MR cavities generally overlap with one end of this correlation (at $n_{13-26} \sim 0$) while the MP cavities are generally consistent with decreasing trends with $n_{13-26}$.
The disk of PDS~70 appears as an outlier in all of these plots for its low $n_{13-26}$; the inner dust disk in this system is known to be variable \citep[][and Appendix \ref{app: IRindex}]{perotti23,gaidos24,Jang2024} and it will be discussed in Section \ref{sec: pds70}.

\subsection{Water spectra in disk cavities} \label{sec: water}
Focusing now on the excitation of water spectra, disks with cavities show properties that reflect the decrease in the high-energy lines, a decrease in column density, and the dominance of colder water. These properties are illustrated in Figure \ref{fig: water_diagrams} with the diagnostic diagrams introduced in \cite{banz25}, where line ratios at different $E_u$ are sensitive to the relative emitting area and column density of water reservoirs across a temperature gradient from the inner dust rim out to the snowline (for a simple interpretation of each diagnostic, see labels in Figure \ref{fig: water_diagrams}). The MR/MP dichotomy of inner disk cavities defined in the previous section and Figure \ref{fig: lum_correl} reveals how they evolve in comparison to full disks in these diagrams. MR cavities have the 6000~K \ch{H2O} line luminosity (which can be interpreted as mostly driven by the size of the emitting region) comparable with those measured in full disks with relatively small areas, while MP cavities are systematically lower than most of the full disks (left plot in Figure \ref{fig: water_diagrams}). The luminosity of the 1500~K \ch{H2O} lines are instead more similar in MR and MP cavities, as already shown in Figure \ref{fig: lum_correl}. 
Cavity disks in general show a higher 3340~K line ratio, where detected, that is consistent with a factor of $\gtrsim 10$ lower column density of the warm water component in comparison to the majority of full disks (middle plot). The only exception among the cavity disks is PDS~70, which has the lowest value of the 3340~K line ratio consistent with the higher column density measured in most full disks; this will be discussed in Section \ref{sec: pds70}. 

\begin{figure*}
\centering
\includegraphics[width=1\textwidth]{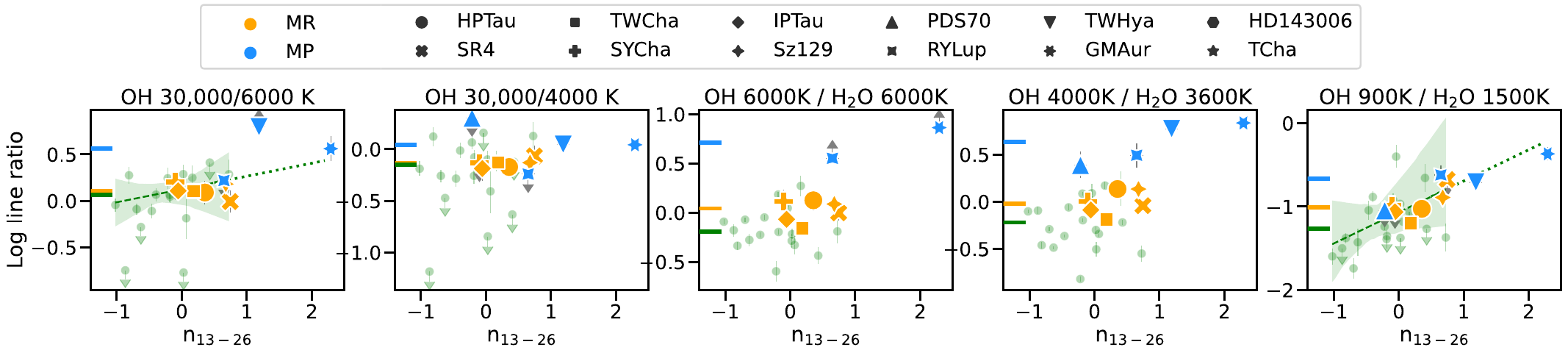} 
\caption{Same as Figure \ref{fig: lum_correl_IRindex} but showing line flux ratios between OH lines or OH and \ch{H2O} lines over a range of upper level energy. Higher median OH/\ch{H2O} ratios are observed in all cavity disks (including MR) in comparison to full disks across energy levels, consistent with a higher \ch{H2O} photodissociation.}
\label{fig: cavities_OH_IRindex}
\end{figure*}

Overall, the sample of cavity disks sits higher-right in the 3600/6000~K and 1500/3600~K diagram in comparison to the full disks, indicative of water spectra being increasingly dominated by colder emission in cavity disks (right plot in Figure \ref{fig: water_diagrams}). The MR cavities are consistent with a similar temperature gradient as in full disks but tend to have a less steep temperature slope \citep[assuming $T = T_0 ( r / 0.5 \text{au})^{-\alpha}$ as in][]{munozromero24a,banz25} and have higher 1500/3600 K ratios indicative of larger cold water areas \citep[right plot, compare to full disks in Figure 22 in][]{banz25}. MP cavities sit at the extreme right of the plot due to their much lower hot water luminosity, indicating even shallower temperature gradients that may in some cases be well reproduced by a single warm ($\sim 400$~K) component \citep[similar to the case of MY~Lup in][]{Salyk25}. The 3600/6000~K and 1500/3600~K ratio values measured in MP cavities are consistent with two disks identified in \citet{Temmink2025} as being ``water-poor" (DN~Tau and CX~Tau). The authors proposed those disks could have a small inner cavity; CX~Tau, with $n_{13-26} \sim 0$ as measured in MIRI, is indeed a good candidate for an unseen small inner dust cavity, while DN~Tau, with $n_{13-26} \sim -0.16$, less so especially without evidence for a mm cavity at high angular resolution \citep{long19}.

In terms of the coldest water detected with MIRI, the diagnostic line ratios in Figure \ref{fig: cold_water_diagr} show that cavity disks have values overlapping to those in full disks (similar to what found in the diagram in Figure \ref{fig: water_diagrams}) but overall they tend to have higher 1500/6000 K ratios as well as higher 1448/1615 K ratios, consistent with 170--200~K water in all disks. This diagram indicates that as the hot water reservoir is decreased in a disk cavity, the spectrum is dominated by colder water near the snowline (see more in Section \ref{sec: disc}).

\subsection{OH emission and \ch{H2O} photodissociation} \label{sec: OH}
Figures \ref{fig: lum_correl} and \ref{fig: lum_correl_IRindex} show that OH emission in all cavity disks tends to be as luminous or slightly super-luminous in comparison to trends defined by full disks as a function of $L_{\rm{acc}}$ or $n_{13-26}$, in contrast to other molecules that show a luminosity decrease at least in MP cavities (if not even in MR cavities, as the case of organics). Due to its efficient re-formation, OH is one of the most persistent molecules that is often a tracer of UV irradiated environments and of water photodissociation \citep[e.g.][]{tappe08,Fedele2013,Parikka2017}. Thanks to the large range of energy levels covered by MIRI spectra, it is now possible to investigate different gas reservoirs because OH is expected to trace a similarly extended radial gradient to water in inner disks \citep[a surface layer higher up in the disk atmosphere, e.g.][]{woitke18,Woitke24} as well as the highly-excited OH lines from photodissociation of water \citep{tappe08,carr14,tabone21}. Specifically, levels of $> 20,000$~K should unambiguously be populated by photodissociation of water showing the characteristic flux asymmetry in OH doublets \citep[e.g.][]{Neufeld2024}, sometimes detected down to levels of $\sim 10,000$~K \citep[in the MP disk of GM~Aur,][]{Romero-Mirza25}, while lower energy levels emitting from longer wavelengths could be increasingly populated by chemical pumping \citep{zannese24,tabone24}. 

The 30,000~K lines are equal to or super-luminous in cavity disks in comparison to the trend defined by full disks. The flux of these lines is proportional to the amount of \ch{H2O} photodissociated per unit of time, not to the OH abundance, and may simply trace a UV irradiated surface layer where OH is abundant and \ch{H2O} under-abundant where \ch{H2O} is constantly photodissociated back to OH \citep{tabone21,zannese24}. The ratio of these very high-energy lines to the 6000~K and 4000~K lines, which are likely excited by chemical pumping, are higher in MP disks and show some correlation with $n_{13-26}$ (Figure \ref{fig: cavities_OH_IRindex}). 
Significant trends are also observed in the OH/\ch{H2O} line ratios with similar upper level energy, which might trace a radial gradient of these two molecules from a hotter inner region to a colder region at larger radii. The OH/\ch{H2O} ratios are on average higher in cavity disks in comparison to full disks across all energy levels included in MIRI (Figure \ref{fig: cavities_OH_IRindex}), possibly indicating increased water photodissociation as the inner disk is becoming depleted from the shielding dust.
A significant correlation is detected in the $\sim 4000$~K OH/\ch{H2O} line ratio in the MP disks, and the colder OH lines covered by MIRI ($\sim 900$~K at 24.6~$\mu$m, the most likely tracer of a collisionally-populated reservoir) show a correlation with $n_{13-26}$ in all disks (Figure \ref{fig: cavities_OH_IRindex}). These trends may be due to the different response to UV irradiation and dust removal, causing the dissociation and recession of water to larger radii while OH persists. Given the complexity of non-LTE OH excitation, all these trends should be investigated in future thermo-chemical modeling work to clarify their origin.

\begin{figure*}
\centering
\includegraphics[width=1\textwidth]{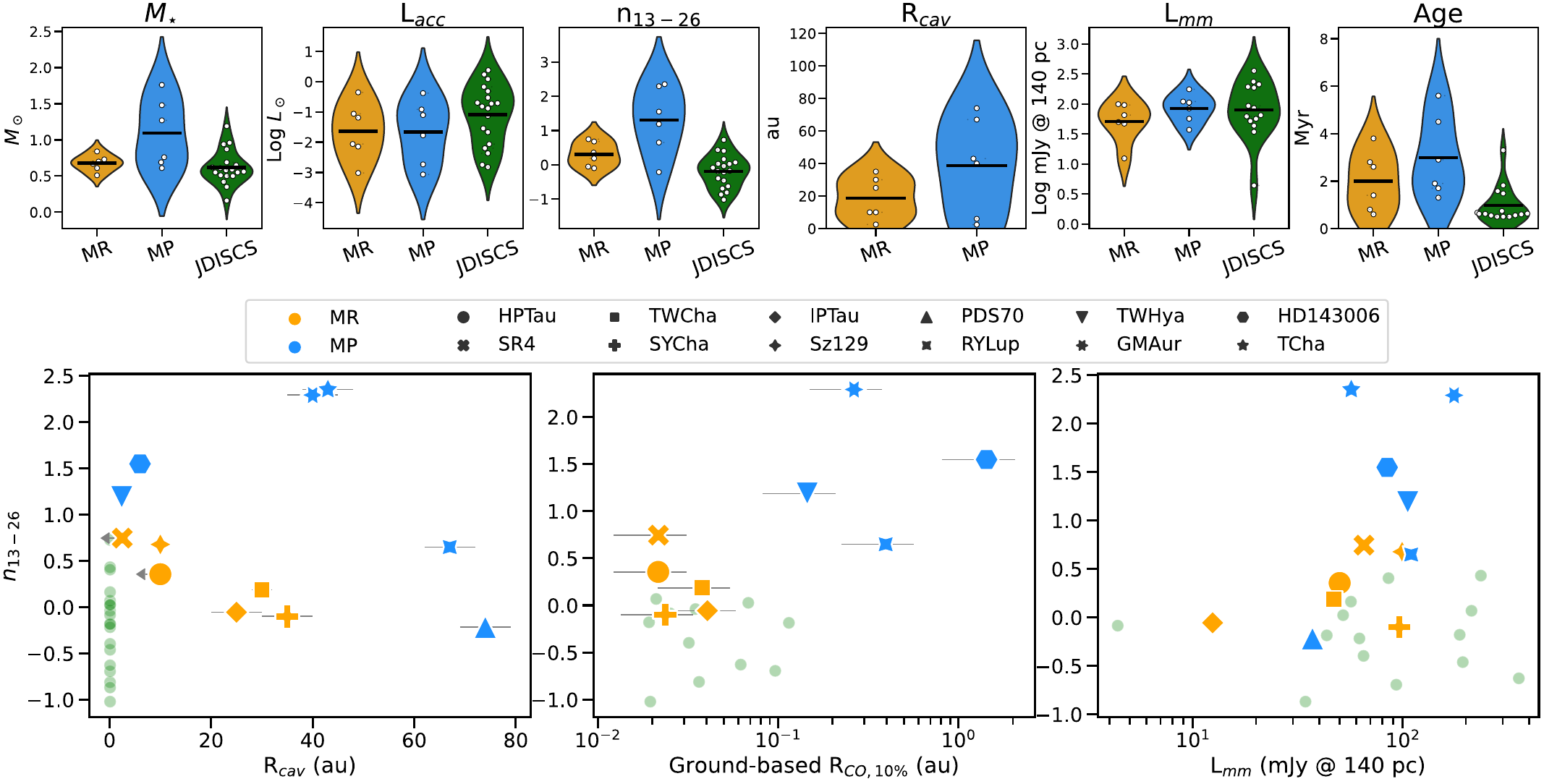} 
\caption{Sample properties for disks with cavities included in this work. Disks with cavities are separated into ``molecule-rich" (MR, orange) and ``molecule-poor" (MP, blue) as explained in Section \ref{sec: analysis}. The reference sample shown in green is the JDISCS-C1 sample (Section \ref{sec: analysis_1}), assuming R$_{cav} = 0$. R$_{cav}$ is the millimeter dust emission cavity size. Top row: violin diagrams showing individual disks as white dots in each distribution. Bottom row: R$_{CO, 10\%}$ is the Keplerian radius of near-infrared spectrally-resolved CO emission at 10\% of the line peak, tracing high-velocity molecular gas from ground-based observations (Table \ref{tab: sample}).}
\label{fig: sample_props}
\end{figure*}

\section{Discussion} \label{sec: disc}
With the improved line luminosity measurements from MIRI spectra (in sensitivity and de-blending, compared to Spitzer) and simultaneous accretion luminosity estimated from mid-infrared HI lines, the new MIRI-MRS spectra expand on previous Spitzer and ground-based results and demonstrate a dichotomy in molecular emission observed from disks that have a millimeter dust cavity (Section \ref{sec: analysis}). Disk cavities are either very similar to full disks from the point of view of their infrared molecular emission (named molecule-rich in this work, or MR), though they still tend to show sub-luminous emission from organics and in particular from \ch{C2H2}, or they are sub-luminous in all the molecular emission except for OH lines in some cases (named molecule-poor or MP). In this section, we describe the properties emerging from the sample included in this work for the two types of disk cavities and discuss their potential interpretations in terms of inner disk evolution.

\subsection{Sample properties of MR vs MP disk cavities} \label{sec: MR_vs_MP_props}
In Figure \ref{fig: sample_props}, we present the emerging properties for the dichotomy into molecule-rich and molecule-poor cavities as defined in this work. The sample includes cavities as small as $\sim 2$~au and as large as $70$~au, with IR index between $-0.2$ (indicating residual inner hot dust within the mm cavity) and up to $\sim 2.4$ (indicating significant dust depletion within the mm cavity). MR cavities have IR index in the range $-0.2 \lesssim n_{13-26} \lesssim 0.75$ with a trend of decreasing $n_{13-26}$ for increasing millimeter cavity sizes at least up to 40~au (bottom left panel in Figure \ref{fig: sample_props}). MP cavities, on the contrary, have $n_{13-26} \gtrsim 1$ with increasing value for cavities up to $\sim 40$~au, and then decreasing $n_{13-26}$ value up to $\sim 70$~au. The discovery of this bifurcation into these two global trends is remarkable, since the index $n_{13-26}$ should reflect more the radial distribution of (sub-)$\mu$m dust rather than the truncation of mm grains outside a cavity (see also Appendix \ref{app: IRindex}). This bifurcation is one of the fundamental findings of this work supporting a dichotomy in molecular emission, as we will discuss below (we will also show in Section \ref{sec: expanding} that the bifurcation of MR and MP in this parameter space is supported by increasing the sample). 

In this sample of T~Tauri stars (we do not include any $T_{\rm{eff}} > 5500$~K in this work), MP cavities equally split between sub-solar and super-solar stellar masses, while MR cavities only have sub-solar masses (Figure \ref{fig: sample_props}). This supports previous findings that stars earlier than K5  more frequently have mm+IR cavity while stars later than K5 more frequently have a mm cavity that does not show a NIR deficit \citep[][]{vanderMarel23}.
The two cavity types, however, are not separated in accretion luminosity and show a $L_{\rm{acc}}$ range similar to that of full disks, though with a lower mean, supporting in this regard previous findings about lower accretion in disk cavities in general (Section \ref{sec: intro}). 
The mm luminosity is comparable in both MR and MP and the full disk sample with no clear trend with $n_{13-26}$, suggesting that the observed mass in mm grains may be similar or slightly larger in MP cavities \citep[for the caveats of converting single-frequency millimeter continuum flux into an estimate of disk solid mass see e.g.][]{miotello2023}. The age spread of MR and MP is also similar, but MP include older objects and the mean age difference between MP and MR is $\sim 1$~Myr in this sample (lower for MR); a potential decrease in molecular emission with time has recently been proposed in \citet{Romero-Mirza25}, where GM~Aur and other disks classified as MP in this work were proposed to be older than SY~Cha\footnote{While \citet{Romero-Mirza25} adopted ages from different literature works, by using the tool by \citet{Deng2025} in this work we actually find a younger age for GM~Aur and older for SY~Cha; individual ages are uncertain \citep{soderblom14}.}, classified MR in this work. The cavity disks in this sample are instead, on average, definitely older than the JDISCS-C1 full disk sample, where ages are obtained in the same way and adopted from Zhang et al. in prep.

In reference to previous findings from ro-vibrational CO lines as velocity-resolved from the ground, which showed a recession of CO and \ch{H2O} to larger radii in disk cavities \citep[][and Figure \ref{fig: water_seq_depl}]{banz17}, we now find that the CO inner emitting radius shows a clear separation between MR cavities, which have broad CO emission extending well into 0.01--0.1~au (fully consistent with where CO is observed in full disks), and MP cavities, where instead high-velocity CO emission is not observed, indicating that CO gas has receded to $>$~0.1--1~au (bottom middle panel in Figure \ref{fig: sample_props}). This clear separation in CO emitting radius shows that the molecular dichotomy observed in disk cavities in this work has a radial evolution component to it, where molecules in MP cavities must be strongly reduced in an inner region that is instead molecule-rich in full disks and MR cavities. 

To summarize, the MR/MP molecular dichotomy in dust cavities does not seem to be primarily due to large differences in stellar or accretion luminosity, nor distinguished by the size of the mm cavity or the disk dust mass (if proportional to the mm luminosity). Bearing in mind the relatively small sample size, the main differences emerging between MR and MP in terms of any properties other than the MIRI molecular emission analyzed in this work are (Table \ref{tab: sample} and Figure \ref{fig: sample_props}): \\
\begin{enumerate}
    \item MR are confined within $-0.2 \lesssim n_{13-26} \lesssim 0.75$ and are a mix of IR, mm, and mm+IR dust cavities, while MP typically have $n_{13-26} > 0$ and they are all mm+IR dust cavities (with the exception of PDS~70); 
    \item the velocity-resolved CO lines indicate narrower lines in MP, showing that the higher velocity, hotter, inner molecular gas has been depleted;
    \item MP include (but are not limited to) super-solar mass stars and larger mm dust cavities;
    \item MP cavities may also be on average older than MR cavities, with mean ages of $\sim 3$~Myr versus $\sim 2$~Myr respectively.
\end{enumerate} 
We also note that the two cavity types are not distinguished by the prominence nor the shape of the 10~$\mu$m silicate feature (see Appendix \ref{app: sample_figures}). This feature is being analyzed in a sample that partly overlaps with this work in Romero-Mirza et al. in prep., where the non-detection of water across the entire MIRI spectrum is found in a few disks that consistently show a very weak silicate feature. This is consistent with what is observed in this sample in the only two disks where \ch{H2O} is not detected at any wavelength (HD~143006 and T~Cha), which may represent the most extreme cases of MP cavities with the least residual inner disk dust within the mm cavity.

\subsection{Dust cavity types and dust filtration} \label{sec: cavity struct}

The lower infrared indices in MR cavities (which extend into slightly negative values consistent with full disks), together with the velocity-resolved CO lines demonstrating molecular gas at $< 0.1$~au, indicate that MR cavities have substantial residual sub-$\mu$m dust and molecular gas in an inner region that is instead more depleted in MP cavities. Building on suggestions made in previous analyses of infrared spectra \citep[][and Section \ref{sec: intro}]{najita10,Salyk15,salyk19,banz17}, these results support a key role for dust filtration through a mm cavity \citep[e.g.][]{rice2006,zhu2012} in sustaining a rich molecular gas chemistry in the planet-forming region. (Sub-)$\mu$m-size dust flowing into a mm cavity through the cavity edge, where instead mm grains are trapped, should in fact sustain molecule formation from \ch{H2}, which most easily forms on dust grains \citep[e.g.][]{glassgold09}, as well as shield them from photodissociation \citep{bethell09,bruderer13}. The role of dust in sustaining molecule formation as well as in shielding them from UV dissociation most likely explains how MR cavities retain spectra that have molecular luminosity more similar to full disks than what is observed in MP cavities. 

The residual dust inside the mm cavity could be in the form of a radially narrow dust belt with a gap between the mm cavity edge at larger radii \citep[the case of so-called ``pre-transitional" disks][see also Appendix \ref{app: additional}]{espaillat07} or diffuse within the cavity perhaps as streamers \citep{dodson11}, geometries that may vary in different disks and are only in some cases directly resolved \citep[e.g. the case of the inner dust belt in DoAr~44,][]{Bouvier20}. Regardless of the specific geometry, the general emerging picture is that mm cavities that still retain enough inner dust to have a similar infrared slope to full disks are of MR type while mm+IR cavities are of MP type, with some exceptions. The case of MR cavities in particular seems to show a mix of situations with IR-only, mm-only, and mm+IR cavities in this sample (Table \ref{tab: sample}), while the only exception in MP cavities is PDS~70 (discussed in Appendix \ref{app: additional}). Future samples will be fundamental to confirm the distribution in each type (see Section \ref{sec: expanding}), and dust modeling work should be done to understand the bifurcation of dust cavities in Figure \ref{fig: sample_props}.

We remark that the weaker line luminosity measured in MP cavities argues against dust evolution simply through growth and settling in these disks. Dust settling is expected to increase the line luminosity of all molecules by removing small dust grains from the disk surface over time, lowering the dust opacity, and therefore increasing the visible column of molecular gas \citep{Greenwood19} that can, at high enough densities, self-shield from UV dissociation \citep[e.g.][]{bethell09,bruderer13,Kanwar25b}. At least some gas density depletion seems therefore to be necessary to explain at least the MP cavities, as we will discuss more in the next sections.
The correlations between molecular luminosities and $n_{13-26}$ in Figure \ref{fig: lum_correl_IRindex}, instead, seem to support the role of dust settling in the atmosphere of full disks: if more negative $n_{13-26}$ values are truly driven by higher levels of dust settling \citep{DAlessio2006,furlan09}, this could explain the increase in line luminosity as explained above based on \citet{Greenwood19}. The range in line luminosity in full disks may therefore be regulated by dust settling producing $-1 \lesssim n_{13-26} \lesssim 0$, while $n_{13-26} \gtrsim 0$ may predominantly reflect dust depletion in the form of a cavity. New dedicated dust model grids are necessary to test the interpretation of all these trends.

\begin{figure*}
\centering
\includegraphics[width=0.8\textwidth]{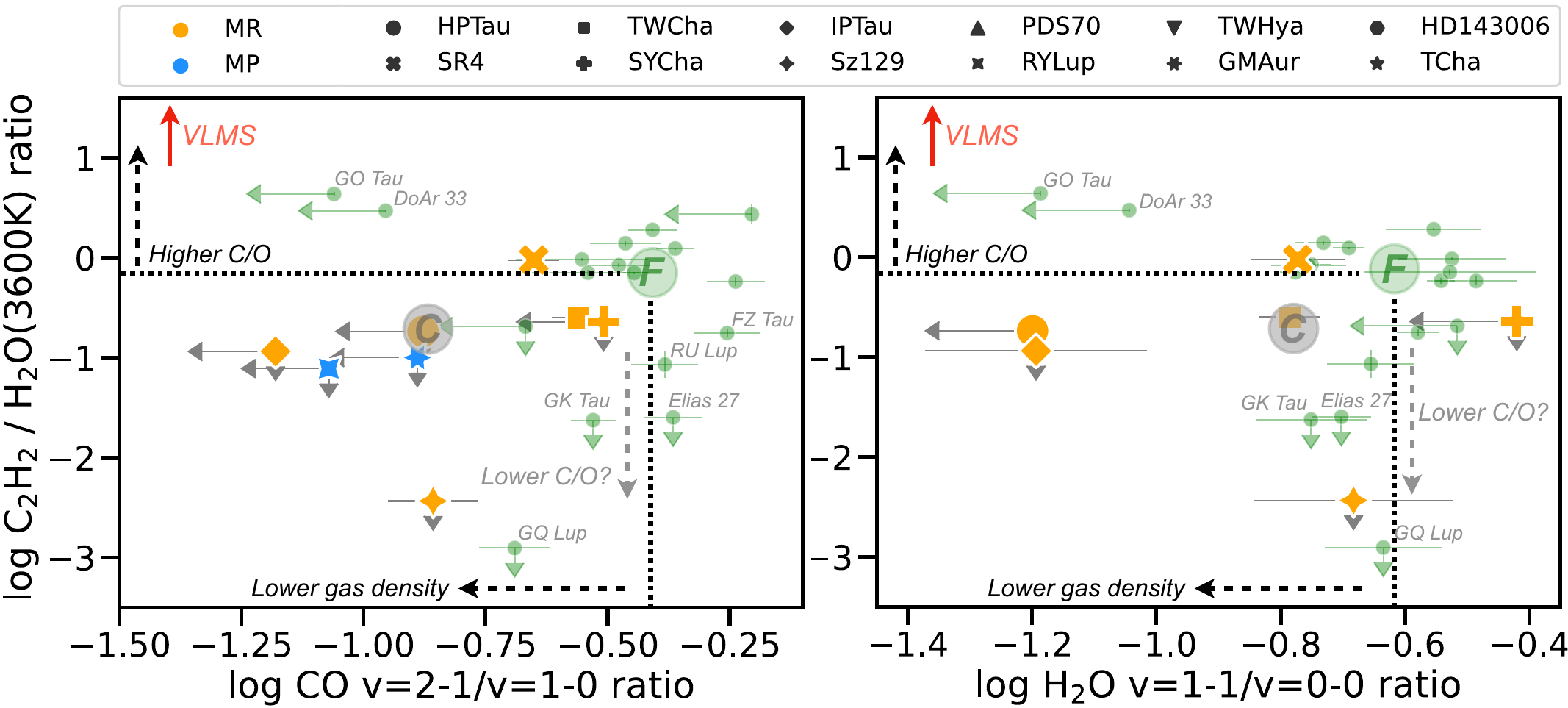} 
\caption{Line flux ratio diagram for \ch{C2H2}/\ch{H2O}, sensitive to the elemental C/O ratio, and the vibrational ratios for CO (left) and \ch{H2O} (right), sensitive to the gas density (see Section \ref{sec: gas density} for a discussion of their interpretation). Median values are shown by the larger circle marked with ``$C$" for cavities and ``$F$" for full disks; the latter is used for reference to identify the suggested interpretation of this diagram in terms of a higher C/O (to the top) versus a lower gas density (to the left). Some T~Tauri disks are consistent with a lower gas density and yet show a higher \ch{C2H2}/\ch{H2O} ratio, DoAr~33 and GO~Tau; thermo-chemical modeling found evidence for an increased C/O ratio in at least one of them \citep{Colmenares24}. The red arrow shows the region of very low mass stars (VLMS), which are found to have even higher \ch{C2H2}/\ch{H2O} ratios as indicative of their super-solar C/O ratio \citep{Arabhavi25,grant25}. Other disks at the opposite corner of the diagram might be due to a lower C/O ratio in the emitting layer (see text for details).}
\label{fig: C2H2_density_drop}
\end{figure*}

\subsection{Gas density in disk cavities} \label{sec: gas density}

While molecular spectra of MP cavities are clearly sub-luminous, even those of MR cavities show a few remarkable differences from those of full disks. Organic emission, particularly from \ch{C2H2}, is overall sub-luminous (Figure \ref{fig: lum_correl}). A similar drop in luminosity is observed in the CO $v$=2-1 lines and, to a lesser extent, in the \ch{H2O} $v$=1-1 lines (Figure \ref{fig: lum_correl}). Moreover, a line ratio identified in previous work as being sensitive to the \ch{H2O} column density in the warm-hot reservoir \citep{banz25,Gasman2025} shows values consistent with a decreased column density in the emitting layer, by a factor of 10 (or more) than in full disks (middle plot in Figure \ref{fig: water_diagrams}). To support that, the OH/\ch{H2O} ratio increases in MR cavities in comparison to full disks (Figure \ref{fig: cavities_OH_IRindex}), which could be due to an increase in \ch{H2O} photodissociation by UV. A lower column density of \ch{H2O} provides a potential important clue for the interpretation for the sub-luminous \ch{C2H2} emission in terms of a lower gas density in the emitting layer, as we will discuss in the following.

To further investigate any evidence for a lower gas density, we consider two additional independent tracers that are covered in MIRI spectra: the excitation of higher vibrational lines of CO $v$=2-1 and \ch{H2O} $v$=1-1 lines relative to their lower vibrational levels, which due to the different critical densities (higher in higher-vibrational levels) are sensitive to the gas volume density in the emitting layer. This has been previously discussed in thermo-chemical models of non-LTE excitation of ro-vibrational CO lines observed in disks \citep[e.g.][]{thi13,bruderer13,Kanwar25b}, finding that CO is generally excited in non-LTE and that the CO $v$=2-1/$v$=1-0 ratio decreases with both the CO volume density and column density due to an increasing deviation from LTE in lower-density gas \citep[Figure 4 in][]{bosman19}. For water, a similar effect is expected in the relative excitation of lines from $v=1$ in comparison to $v=0$, where the sub-thermal excitation of the former is due to deviation from LTE as the gas density decreases from the critical density of these transitions \citep{meijerink09,bosman22}. Both the $v=1-0$ and $v$=1-1 lines of water observed with MIRI have been confirmed to be excited in non-LTE, suggesting a gas density of $<< 10^{13}$~cm$^{-3}$ in the emitting layer \citep[see Figure 7 and its discussion in][]{banz25}. 

We show the CO $v$=2-1/$v$=1-0 and \ch{H2O} $v$=1-1/$v$=0-0 ratios measured in this work in Figure \ref{fig: C2H2_density_drop}. The lines used in the ratios are the same as shown in Figure \ref{fig: lum_correl}, where for the \ch{H2O} $v=0-0$ line we use the hot component that has similar upper level energy as the \ch{H2O} $v$=1-1 line ($E_u = 6000$~K). Both these ratios on average decrease in MR cavities in comparison to full disks, suggesting that indeed the gas density is lower and supporting the lower column density suggested by the $v=0-0$ line ratio in Figure \ref{fig: water_diagrams}. The deviation between full disks and cavity disks is much more prominent in the CO ratio than in the \ch{H2O} ratio; while the MR cavity sample needs to be increased to better assess its distribution, a larger deviation from LTE excitation in the CO $v$=2-1 lines may be due to their higher critical density \citep[$\gtrsim 10^{13}$~cm$^{-3}$][]{thi13,woitke16,Kanwar25b}. 

Three independent lines of evidence in MIRI spectra therefore point towards a decreased gas density in the emitting layer in disk cavities: the flux ratio of \ch{H2O} $v=0-0$ lines that have different Einstein-$A$ coefficient, and the sub-thermal excitation of \ch{CO} $v$=2-1 lines and, to a lesser extent, the \ch{H2O} $v$=1-1 lines. Confirmation of this interpretation for the line ratios requires dedicated thermo-chemical modeling applied to inner disk dust cavities, which so far has mostly been applied to the pure rotational lines of CO \citep{bruderer13}.
We remark that a lower gas surface density may already be implied by the lower accretion rates measured in disk cavities \citep[Figures \ref{fig: sample_props} and \ref{fig: Manara_compar}, and][]{Najita2007_TD}, which in this case would imply that the lower density is not just in the emitting layer observed with MIRI but in the accretion region and possibly the inner disk itself. A lower emitting gas mass in inner disk cavities in comparison to full disks has indeed been found from fits to ro-vibrational CO spectra \citep{salyk11,banz15} and to rotational \ch{H2O} emission (Romero-Mirza et al., in prep.).
A lower density could also provide an explanation for the more prominent cold water emission in cavity disks on average in comparison to full disks, both in the 1500/3600~K and 1500/6000~K ratios and in the 1448/1615~K ratio (Figures \ref{fig: water_diagrams} and \ref{fig: cold_water_diagr}). While an increased pebble drift is proposed to explain increased cold emission in full disks \citep[][]{banz23b,krijt25}, a decreased hot-warm gas density may expose more prominently the colder emission closer to the midplane snowline in cavity disks, especially in MP cavities. Another potential explanation is that the water spectrum in disk cavities may simply look colder due to a larger deviation from LTE \citep{Vlasblom25}, which again could be due to a lower gas density.

The other prominent difference between cavity disks and full disks is that organic emission, especially from \ch{C2H2}, is on average sub-luminous not only in MP cavities but also in MR cavities (Figures \ref{fig: lum_correl}). In Figure \ref{fig: C2H2_density_drop} we now show the \ch{C2H2}/\ch{H2O} ratio, which has become a leading proxy for the elemental C/O ratio as observed with MIRI in inner disks \citep{tabone23,Colmenares24,Arabhavi25,grant25,long25}, even if these molecules do not necessarily trace the same disk region, especially at high C/O ratios, and therefore this molecular ratio does not directly trace just the elemental C/O ratio \citep[see][]{Kanwar25b}. In the ratio, we use the intermediate energy levels for water to include a few MP cavities in the diagram, those where both CO and water are detected, but we remark that the relative distribution of the other points in the diagram does not significantly change if we use the higher-energy lines of water. 
Thermo-chemical models show that the formation of organic molecules is slower than \ch{H2O} and it is driven by X-ray chemistry \citep[e.g.][]{walsh15,anderson21,Woitke24}. Contrary to \ch{H2O}, organic emission should become weaker with increasing UV irradiation due to photodissociation that counters the X-ray induced production \citep{kamp17,Woitke24}, in agreement with the disappearance of organic emission during strong UV outbursts \citep{banz12}. If MR cavities have an increased UV penetration into the inner disk due to reduced dust opacity in comparison to full disks, enhanced UV dissociation of organics could in principle explain their weaker emission even if, at high densities, they are expected to self-shield \citep{Woitke24,Kanwar25b}. However, recent models also propose that molecular shielding from dominant oxygen carriers in the upper disk layers should be fundamental to provide additional UV shielding to organics. Some models focused on water in particular, finding the most dramatic difference in \ch{C2H2} that should be efficiently destroyed and show little to no infrared emission when water shielding from UV is absent \citep{duval2022,Kanwar2024}. This is strikingly in agreement with the drop estimated in water column density in MR cavities from $\approx 10^{18}$ to $\approx 10^{17}$~cm$^{-2}$ (Figure \ref{fig: water_diagrams}, middle plot), which is below the efficient UV shielding limit \citep{bethell09}. It is also in agreement with the gas density drop suggested by the sub-thermal excitation of CO $v$=2-1 and \ch{H2O} $v$=1-1 lines, as discussed above. The gas density is also important for molecule formation, and a density decrease would further limit the formation of organics such as HCN and \ch{C2H2} to counter their dissociation.

The data in Figure \ref{fig: C2H2_density_drop} also suggests interesting implications for the elemental C/O ratio in the gas, which strongly affects the relative emission between organics and water \citep[e.g.][]{najita11,woitke18,anderson21,Kanwar25b}. The C/O ratio probably does not increase in disk cavities in comparison to full disks. This can be concluded from comparison to two full disks that show a decreased gas density and yet an unusually strong \ch{C2H2} feature relative to other molecules: DoAr~33 and GO~Tau \citep[their remarkably similar \ch{C2H2} spectra can be seen in][]{Arulanantham25}. The former was analyzed with thermo-chemical models in \citet{Colmenares24}, suggesting that a much higher C/O ratio of 2--4 is necessary to reproduce the strong organic emission relative to water, similarly to what is proposed for VLMS disks that typically show an even higher \ch{C2H2}/\ch{H2O} ratio \citep{Arabhavi25,grant25,Kanwar25b}. At the opposite corner of the diagram, a sub-group of full disks seem to be consistent with normal or only slightly reduced gas density and a lower C/O ratio; some of these disks had dubious broad features under the organics that have been attributed to solid state \citep{Arulanantham25}. It is interesting that two of these disks, GQ~Lup and GK~Tau, have prominent cold water emission previously attributed to oxygen-enrichment by pebble drift \citep{banz25,Romero-Mirza25,krijt25,Vlasblom25}.
The diagram in Figure \ref{fig: C2H2_density_drop} may become useful in the future to study the effects of gas density versus C/O ratio to explain organic emission in inner disks in general, not only in disks with a dust cavity. The suggested interpretation of the diagram axes as tracing the C/O ratio and gas density in the emitting layer discussed in this section should be thoroughly investigated with future dedicated thermo-chemical modeling applied to full disks and cavity disks, which should assess also the effects of changes in temperature and dust opacity in the observable disk layer.

\begin{figure*}
\centering
\includegraphics[width=1\textwidth]{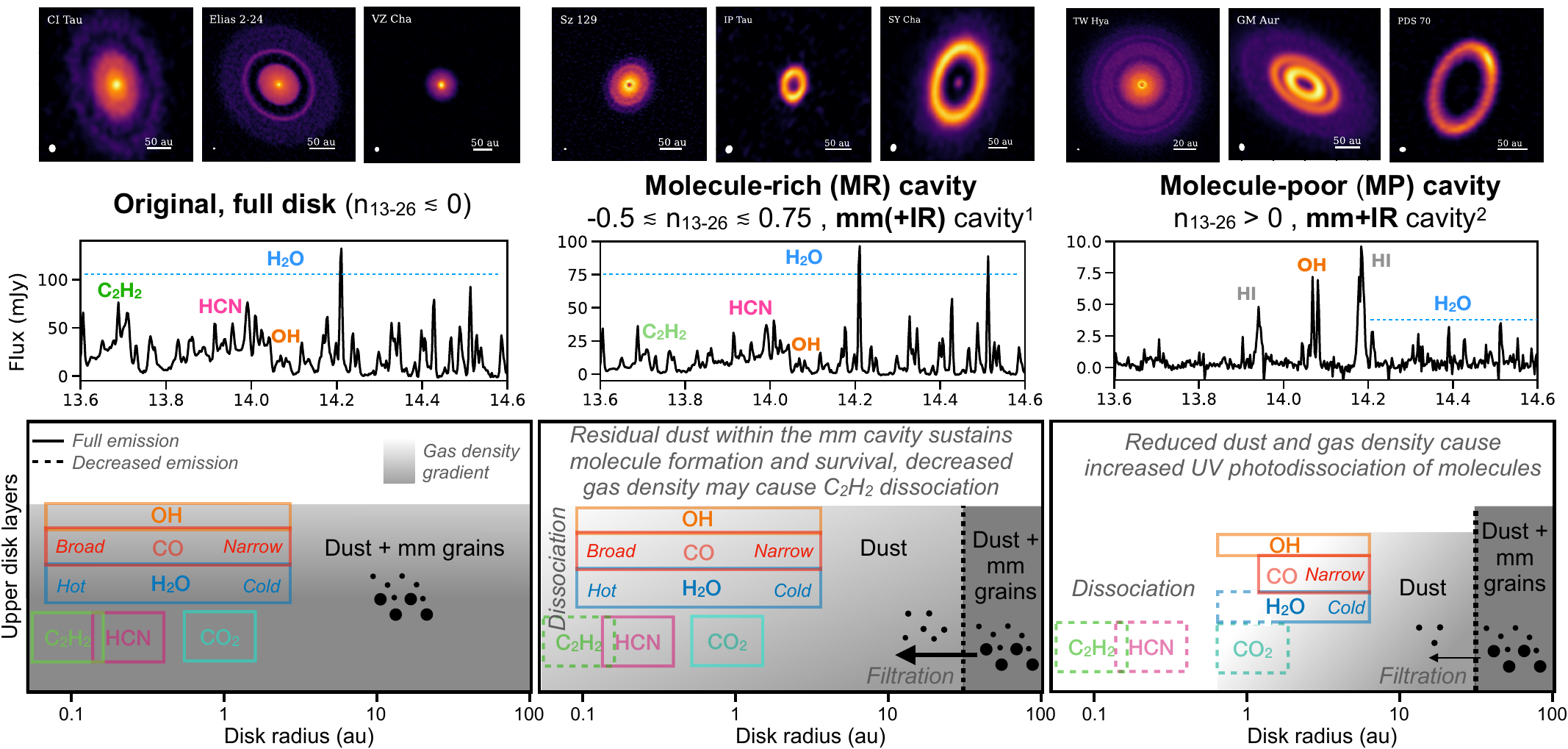} 
\caption{Schematic illustration of the properties and relative distribution of $\mu$m-size dust, mm grains, and molecular gas in the two cavity types presented in this work in comparison to a full disk \citep[with representative molecular emitting regions adopted from][]{Woitke24}. The illustration includes several simplifications and is not meant to reproduce the exact geometry of the inner disk nor which process is opening the mm cavity; see Section \ref{sec: disc} for a discussion of each dust and gas component in the different cavity types. Some gradients observed in the data are labeled for reference, including the temperature gradient in \ch{H2O} and line-broadening gradient in CO. The reduced \ch{C2H2} emission in MR cavities is interpreted as due to a decrease in gas density (Figure \ref{fig: C2H2_density_drop} and Section \ref{sec: gas density}). Representative portions of MIRI spectra at 13.6--14.6~$\mu$m are included above each disk type for reference; spectra for the entire sample are reported in Appendix \ref{app: sample_figures}. ALMA continuum images are reported to the top with examples for each disk type (see Section \ref{sec: obs}). NOTES: $^1$- MR cavities are of mixed dust type but they clearly show a larger IR dust contribution at any mm cavity size (Figure \ref{fig: sample_props}); $^2$- MP cavities are predominantly mm+IR dust cavities, with the exception of PDS~70.}
\label{fig: new_cartoon}
\end{figure*}

\subsection{Dust and gas evolution, and molecule survival} \label{sec: cartoon}

In Figure \ref{fig: new_cartoon}, we provide a simplified illustration to summarize the relative distribution and evolution of dust, mm grains, and molecular gas emerging from this work for MR and MP. For reference, we adopt approximate emitting regions from thermo-chemical models of a full disk that show a radial and vertical stratification of molecules as observed with MIRI \citep[Figure 11 in][]{Woitke24} where organics emit from a narrower and hotter inner region while water, CO, and OH come from a higher and more radially extended layer to account for their observed radial gradients. In the figure, we show a potential interpretation of the approximate evolution of those emitting regions to reconcile all the observational results described above in this work. 

In a MR cavity, dust filtration through the millimeter cavity (indicated by the lower $n_{13-26}$ index relative to an MP cavity, see Figure \ref{fig: sample_props}) may play a key role in sustaining water and CO to similar levels as in full disks, as indicated by the luminosity observed in these molecules (Figure \ref{fig: lum_correl}). 
In a MP cavity, instead, the reduced dust (indicated by the higher $n_{13-26}$ index) may lead to a general dissociation of molecules reducing all the hot gas tracers and leaving only a cold residual \ch{H2O} layer (see Section \ref{sec: water}), narrow CO emission, and a persistent layer of OH emission from water photodissociation (see Section \ref{sec: OH}). The gas density may decrease in all cavities and in particular in MP, as suggested by the decrease in luminosity of CO $v$=2-1 and \ch{H2O} $v$=1-1 lines relative to their lower vibrational lines (Figures \ref{fig: lum_correl} and \ref{fig: C2H2_density_drop}), extending the UV photodissociation of molecules down to deeper layers, which could explain the increased OH/\ch{H2O} ratio and the drop in \ch{C2H2} luminosity in both MR and MP cavities (Figures \ref{fig: cavities_OH_IRindex} and \ref{fig: C2H2_density_drop}), when molecular shielding from the main oxygen carriers in the upper disk layers is reduced.


\subsection{Common evolution or different processes?} \label{sec: origin}
Previous work suggested that disks with spatially-resolved millimeter dust cavities may be a long-lasting evolutionary phase that happens only in more massive disks \citep{merin10,Owen2016,pinilla18,pinilla20,vanderMarel18}, while most disks might evolve to disperse through an ``anemic"/``homologously depleted"/``evolved" phase where the SED decreases at all wavelengths without signs of developing an inner dust cavity first \citep{lada06,hernandez07,currie11,Ingleby11,sic-agu11}. The older age and high mm luminosity of cavity disks in this work (Figure \ref{fig: sample_props}) do support the idea that at least some of these disks are long-lived and as massive as full disks. 
Within this general framework, we wish to inquiry whether the MR/MP dichotomy may be a sign of sequential rather than different evolutionary paths.

\begin{figure}
\centering
\includegraphics[width=0.45\textwidth]{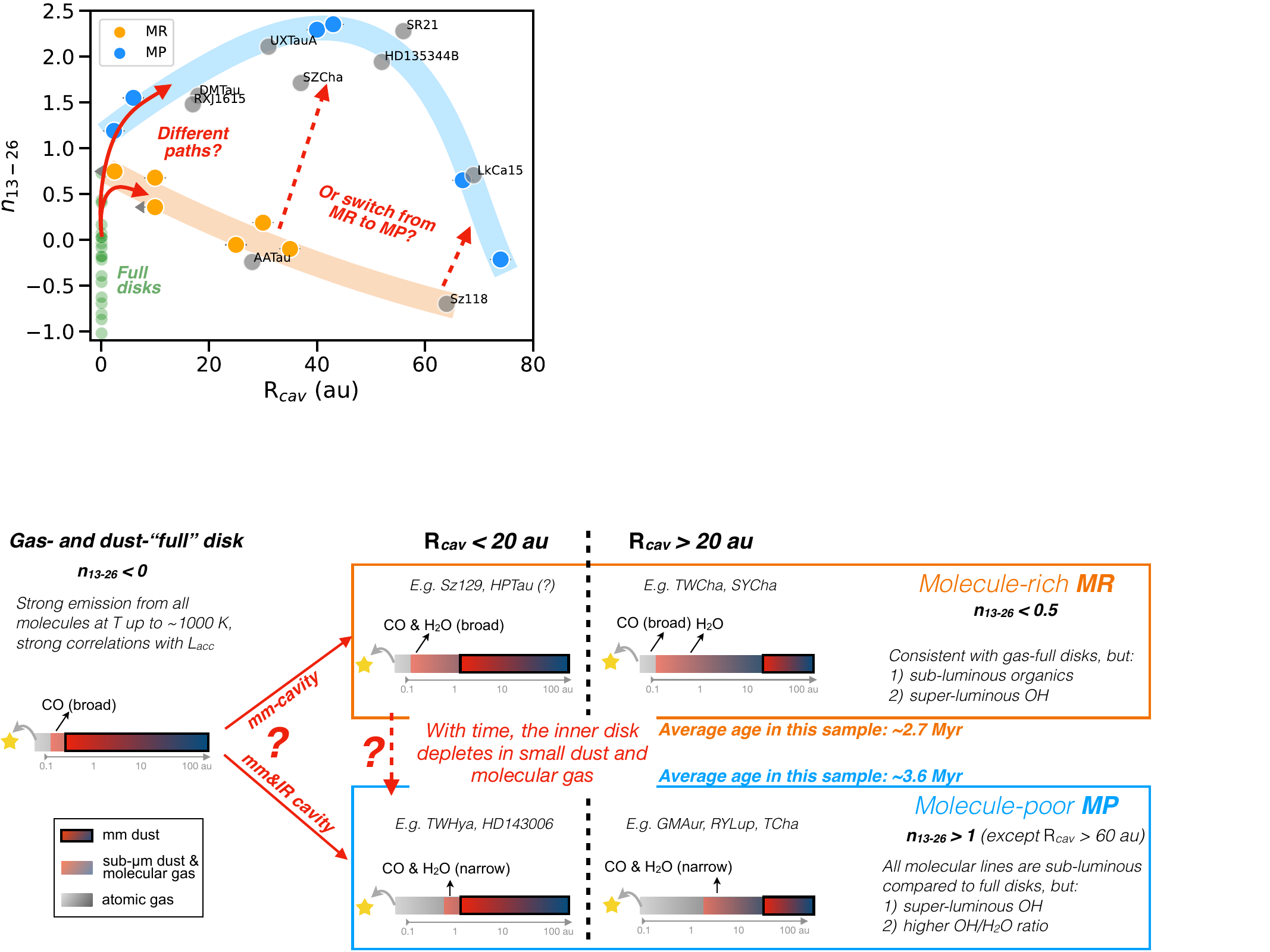} 
\caption{Same as bottom left panel in Figure \ref{fig: sample_props} but marking the two potential interpretations discussed in Figure \ref{fig: new_cartoon} and in the text. The regions defined by the current samples of MR and MP are highlighted by eye with shaded orange and blue bands respectively. The grey datapoints show additional disks observed with MIRI as of Cycle 4 with $n_{13-26}$ measurements available from previous Spitzer spectra.}
\label{fig: prediction}
\end{figure}

As described above, we find that the MR/MP dichotomy does not relate to the millimeter cavity size nor accretion onto the star, but it is instead related a drop in gas density and to the residual content of small ($\lesssim \mu$m) dust grains within the mm cavity. A possible evolutionary scenario to reconcile these aspects is that all cavities may at first form as MR, developing from full disks by opening a mm cavity at any radius (2--70~au, the range observed in this sample). All these mm cavities may at first retain enough $\mu$m-size grains, supplied by filtration through the cavity edge, to sustain the formation and survival of molecular gas inside the cavity. 
As time passes, the $\mu$m-size dust mass within the cavity may decrease by an imbalance between stellar accretion or grain growth and dust filtration through the cavity edge \citep[e.g.][]{zhu2012}. A decrease in inner disk dust in turn increases the observed $n_{13-26}$ index and moves up the disk in Figure \ref{fig: prediction} to eventually become an MP cavity. In the context of this potential scenario, the small age separation (with MP cavities that are on average $\sim 1$~Myr older than MR) is in stark contrast with the net separation of MR and MP in the $n_{13-26}$ vs R$_{cav}$ plane in Figure \ref{fig: sample_props}, suggesting that if it is indeed a time evolution it must be rather fast to switch from MR to MP. Perhaps, due to shielding (dust shielding gas, and molecular shielding from CO and \ch{H2O}) there is a fast tipping threshold between a molecule-rich to a molecule-poor surface layer due to dissociation, a threshold that may depend on a specific dust and gas density. 

To determine the inner disk structure, it is very important to understand what happens to the dust after its filtration through the mm cavity. Using dust evolution models applied to the disk of PDS~70, \citet{pinilla24} show that dust growth to mm grains in the inner 10~au depends on the mass of small dust grains filtering through a pebble trap further out. As the dust growth timescale is inversely proportional to the dust-to-gas ratio \citep[e.g.][]{birnstiel2024}, if the dust-to-gas mass ratio in the inner disk is reduced and the inner gas mass is kept constant or it decreases in much longer timescales, then the grain growth timescales are much longer. If the dust mass in the inner disk drops below ${\sim}10^{-4}~M_\oplus$, re-coagulation becomes too slow and the inner disk contains only ${\lesssim}1-10~\mathrm{\mu m}$ grains \citep[Figs. 1 and 2 in][]{pinilla24}, effectively becoming a mm-cavity disk that does not show an IR deficit. 
Initially, as gas and dust masses are relatively high, it is expected that molecules (e.g., CO, \ch{H2}) can readily survive \citep{bruderer13}, and the mm cavity would be of MR type. However, as the gas and dust content in the inner disk dwindles, another transition may take place. Using thermochemical models of disks with gaps and different depletion levels, \citet{bruderer13} shows that once the gas mass interior to 10~au drops below ${\sim}0.2~M_\oplus$, shielding by residual small dust is needed to allow molecules to survive \citep[Fig. 8 and Table 2 in][]{bruderer13}. Lowering this residual dust mass from ${\sim}2\times10^{-4}M_\oplus$ to ${\sim}2\times10^{-9}M_\oplus$ removes this shielding effect and causes a dramatic decrease in the column densities of CO and \ch{H2} \citep[Fig. 8 in][]{bruderer13}, possibly providing a transition from MR to MP.

In summary, previous modeling of dust cavities, while it still needs to be applied to test a transition between MR and MP, can in principle support a sequence of events in which a molecule-rich mm cavity may form first and the transition from MR to MP takes place later, with both transitions being triggered by their own inner disk dust (and gas) mass thresholds.
This scenario may also explain why cavities around IMTTS (and Herbig) stars are typically MP, if disks around higher mass stars evolve more rapidly and move to the MP phase earlier. 

A possible alternative is that, instead, dust cavities may form from the start as either MR or MP depending on a different cavity-opening process. Given their similar stellar accretion, cavity size, and dust mass as inferred from the mm luminosity (Figure \ref{fig: sample_props}), the different processes giving origin to MR rather than MP should act on a similar radial disk region and enable similar accretion onto the star, but differ in the amount of residual $\mu$m-size dust within the millimeter cavity and possibly in the gas density as discussed above. Multiple processes have been proposed to form a dust cavity: grain growth and settling, planet dynamical interactions, disk winds, dead zones \citep[e.g.][and references therein]{Najita2007_TD,espaillat14,ercolano_pascucci2017,vanderMarel23}. Models including planets of different mass have been particularly successful in explaining bright disks with large cavities that still retain accretion onto the star \citep{Marsh1992,Najita2007_TD,dodson11,huang24}. The combination of planets or dead zones with MHD or photoevaporative winds has been increasingly successful in explaining a large range of conditions up to larger cavities and higher accretion rates \citep{pinilla16,ercolano_pascucci2017,hendler18,garate21,martel2022}. The MR/MP dichotomy provides new observational ground to investigate these scenarios in future work and determine if the dichotomy in inner disk molecular emission is a proxy for a specific cavity-opening process over another. It is worth noting that recent work using GAIA astrometry proposed a $>20$~M$_{jup}$ companion in T~Cha and a $>50$~M$_{jup}$ companion in RY~Lup \citep{vioque25}, which in addition to the known proto-planets in PDS~70 \citep{Keppler18,Haffert19} suggests that at least some MP cavities may be produced by massive companions.

\newpage

\subsection{Expanding the sample} \label{sec: expanding}
Figure \ref{fig: prediction} shows again the dichotomy in MR and MP in the IR index vs mm cavity size space emerged from this work (based on Figure \ref{fig: sample_props}), and marks the bifurcation into two trends discussed above in this section. In this new figure, we include as grey datapoints additional cavity disks that have been observed with MIRI in other programs \citep[][Perotti et al. in prep, Romero-Mirza et al. in prep]{espaillat23,espaillat24} where the IR index can be measured from previous Spitzer data. This larger sample supports the net separation of disk cavities that is emerging from this work in the $n_{13-26}$ vs R$_{cav}$ plane, with a sharp bifurcation into mm+IR cavities in the MP branch and MR cavities that for smaller R$_{cav}$ are mm+IR cavities too but extend into negative values of $n_{13-26}$ at R$_{cav} \gtrsim 30$~au.
Future analyses of larger samples of MIRI spectra will enable testing if the molecular emission follows the two types as predicted here and will better characterize the distribution of properties in the two cavity types, specifically addressing the fundamental questions on their evolution discussed above (including a possible age difference). 

Based on cavity disks that have previous Spitzer data, using the $n_{13-26}$ value we can compile a list of additional disks for each cavity type that could be tested in the future with MIRI spectra (targets with an asterisk have already been observed with MIRI in other programs and are included in Figure \ref{fig: prediction} using their available Spitzer-based $n_{13-26}$ values):
\paragraph{Predicted MR cavities:} AA~Tau$^*$, Sz~118$^*$, CIDA~1, HP~Cha, Sz~100, WSB~60, RY~Tau.
\paragraph{Predicted MP cavities:} DM~Tau$^*$, LkCa~15$^*$, UX~Tau~A$^*$, SZ~Cha$^*$, SR~21$^*$, HD~135344B$^*$, RXJ~1615$^*$, RXJ~1852, CS~Cha, V1247~Ori, LkHa~330, CQ~Tau, MWC~758.\\

\subsection{How can we identify an inner disk cavity?} \label{sec: cavity_identify}

As we have discussed in this work, each observable that has been used so far to identify a dust disk cavity has its own limitations (Section \ref{sec: intro} and Appendix \ref{app: IRindex}). Millimeter cavities can be directly imaged only when larger than the spatial resolution achieved with millimeter interferometers, which in principle means $\sim5$~au but often it is much larger than that. Small ($<15$~au) dust cavities have therefore mostly been missed in the millimeter surveys obtained so far even with ALMA \citep{vanderMarel23}. On the other hand, the infrared index $n_{13-26}$ also does not reveal all the inner disk cavities, which can extend well into the negative values that are typically attributed to full disks (Figure \ref{fig: prediction}). The reason why we can clearly identify some cavities in disks with $n_{13-26} < 0$ in this work is that we have a direct detection of the mm cavity from ALMA. However, would it be possible to identify a disk with a cavity in absence of a direct detection from ALMA? This question is especially relevant in the case of star-forming regions that lack the highest-resolution ALMA data or that are further away, especially at kpc distances where the ALMA resolution decreases significantly. These disks are still well observable at the sensitivity of JWST-MIRI \citep[e.g.][]{ramireztannus2025}, and infrared spectra provide invaluable data to study the evolution of planet-forming regions especially at these distances.

We can use the results of this work to propose specific observables that can be used to identify a disk cavity purely from a MIRI spectrum, assuming an ALMA image is not available or not at spatial resolution better than $\sim 40$~au. These criteria should be valid for T~Tauri disks at $\sim0.5$--5~Myr around solar-mass stars (0.5--1.3~M$_{\odot}$), as included in this work (Table \ref{tab: sample}). First of all, $n_{13-26} > 0$ should be a first indicator of a disk cavity (Figure \ref{fig: sample_props}), particularly when molecular emission is sub-luminous in comparison to trends defined by full disks as a function of accretion luminosity (Figure \ref{fig: lum_correl}). A notable exception can be disks viewed at high inclinations ($\gtrsim 70$~deg), which will decrease the infrared flux at shorter wavelengths and increase the $n_{13-26}$ index similarly to an inner cavity \citep[e.g.][and Section \ref{app: IRindex}]{Ballering19}. However, MR cavities can also have $n_{13-26} < 0$, when enough residual dust is present within the mm cavity as discussed above in this work. The bifurcation discovered in Figure \ref{fig: sample_props} now suggests that future work should explore and determine a more suitable way to classify IR cavities based on their SED. In ambiguous cases, MIRI spectra offer a number of other observables that may reflect the depletion of the inner disk: in this work we have identified in particular that sub-luminous \ch{C2H2} and sub-luminous higher-vibrational-state lines of CO and \ch{H2O} (Figure \ref{fig: C2H2_density_drop}) may provide evidence for a decrease in gas density in the hot inner gas, even in disks where the infrared index alone does not provide clear evidence for dust depletion. This finding indicates a path for future work to study in larger samples and dedicated model explorations the interconnection of dust and gas evolution and the onset of gas depletion in planet-forming regions, with its fundamental implications in planet formation.

\section{Summary \& Conclusions} \label{sec: concl}

Expanding on previous findings from Spitzer and ground-based spectra \citep[][see Section \ref{sec: intro}]{najita10,salyk11_spitz,Salyk15,salyk19,banz17}, in this work we have formally defined two types of T~Tauri disk cavities based on their observed infrared molecular spectra. We compare the MIRI-MRS measured molecular luminosities to those from full disks as a function of accretion luminosity, which has the strongest correlation with the molecular luminosity in full disks \citep[][and Appendix \ref{app: new_lines_4_corr}]{banz20,banz25}. ``Molecule-rich" (MR) cavities have a molecular luminosity that is generally consistent with that measured in full disks with similar accretion, except in the case of organic molecules that can be sub-luminous, while ``molecule-poor" (MP) cavities are significantly sub-luminous in all molecules except in some cases the OH lines.

From the point of view of the solids, these two cavity types are not separated by the millimeter cavity size R$_{cav}$ (which however extend to larger sizes in MP cavities in this sample) nor the millimeter luminosity (and in turn supposedly the mass). However, we discover a bifurcation with two distinct branches in the $n_{13-26}$ vs millimeter R$_{cav}$ plane indicative of different degrees of depletion in (sub)$\mu$m-size dust within the millimeter cavity (Figures \ref{fig: sample_props} and \ref{fig: prediction}): for the same R$_{cav}$, MR cavities have significantly lower $n_{13-26}$ values demonstrating residual dust that can help molecule formation and shielding from UV dissociation. The stark dichotomy and bifurcation in this diagram are fundamental results from this work that are currently supported by increasing the sample with cavity disks from other programs (Figure \ref{fig: prediction}). The infrared $n_{13-26}$ vs millimeter R$_{cav}$ diagram may become an important framework for future studies of the physical and chemical structure of disk cavities and their evolution.  

From the point of view of the gas, MR and MP are not separated in accretion luminosity but the molecular gas shows strong differences (Figure \ref{fig: lum_correl}). MR cavities have several properties consistent with those observed in full disks: broad CO lines (as velocity-resolved from the ground, Figure \ref{fig: sample_props}) indicating emission extending to $< 0.1$~au and strong emission from hot molecular gas tracers (including high-energy \ch{H2O} and CO). However, organic emission tends to be sub-luminous, especially in \ch{C2H2}, and water is consistent with a $\sim 10$ times lower column density and to be dominated by colder emission, confirming and expanding earlier evidence emerged from Spitzer and ground-based spectra (see above). A new discovery of this work is that all T~Tauri disks with a cavity have sub-luminous CO $v$=2-1 and, to a lesser extent, \ch{H2O} $v$=1-1 lines, suggesting further deviation from LTE excitation as due to a decrease in gas volume density in the emitting layer (Figure \ref{fig: C2H2_density_drop}). Together with the on-average lower stellar accretion in comparison to full disks (Figures \ref{fig: Manara_compar} and \ref{fig: sample_props}), multiple lines of evidence therefore suggest a decreased gas density in inner disk cavities, including in MR cavities.
Additionally, MP cavities have narrow CO lines profiles indicating the emission has receded to $> 0.1$--2~au and have sub-luminous emission from all hot gas tracers except OH. Water is not detected in the hotter reservoir typically observed in full disks, and is characterized by steeper temperature profiles dominated by emission at 170--400~K, where detected. Two disks in this sample (T~Cha and HD~143006) show the extreme case of MP cavities with non detection of water at all MIRI wavelengths and weaker 10-$\mu$m silicate emission, consistent with findings from Romero-Mirza et al. in prep. These extreme conditions are probably reached by all MP cavities with time, and may be more common at later stellar ages.

The OH emission stands in stark contrast with other molecules by showing a general persistence in disk cavities in transitions across all upper level energies (900--40,000~K), indicative of a persistent \ch{H2O}-photodissociation layer as well as possibly different reservoirs down to a collisionally-dominated colder layer. Compared to full disks, the measured OH/\ch{H2O} ratios across energy levels increase in MR cavities and even more in MP cavities (Figure \ref{fig: cavities_OH_IRindex}), suggesting an increase in water photodissociation possibly connected to the decrease in gas density (Section \ref{sec: gas density}).
Put together, the infrared molecular emission observed in disk cavities suggests a feedback where lowered dust within a mm cavity enhances water photodissociation from a hotter layer first out to a colder layer later; in turn, water dissociation decreases its column density and may additionally enhance the organic destruction rates beyond that driven by the loss of small dust grains. Molecular shielding by \ch{H2O} may be particularly important for organic survival in addition to dust shielding and could explain the decrease in organic emission in MR cavities that show evidence for a decrease in gas density, especially in the case of \ch{C2H2} as proposed in recent models \citep{duval2022}. This interpretation should be tested with future thermo-chemical modeling studies that explore how molecular abundances may vary in full disks and in inner disk cavities under a range of dust and gas depletion factors and a range of elemental C/O ratios, to reproduce different conditions that the distribution of disks in the diagram in Figure \ref{fig: C2H2_density_drop} may reflect.

As discussed in Section \ref{sec: disc}, the MR/MP dichotomy could be due to different physical conditions in cavity-opening processes that produce different levels of dust and gas depletion early on and lead to the bifurcation in Figure \ref{fig: prediction}, or to a common evolution where cavities first form as MR and then switch to MP. The net separation in the $n_{13-26}$ vs R$_{cav}$ plane suggests that, if part of the same evolution, the switch from MR to MP may happen rapidly and depend on a tipping point across a threshold in dust and gas density. While the limited sample included in this work starts to reveal emerging properties for each cavity type, it will be fundamental to expand the sample to better characterize the typical properties and distribution of each type and their origin in connection to disk evolution and planet formation.

\acknowledgments
The authors acknowledge feedback from an anonymous referee who helped improving significantly the discussion of organics, as well as helpful discussions on the topic with Jayatee Kanwar. The authors are thankful to Marissa Vlasblom for providing new model line fluxes for comparison to the data in Figure \ref{fig: CO2_cavity}, and to Joshua Sendgikoski for help with stellar age estimates.
This work is based on observations made with the NASA/ ESA/CSA James Webb Space Telescope. The JWST data used in this paper were obtained from the Mikulski Archive for Space Telescopes (MAST) at the Space Telescope Science Institute and can be accessed via:
\dataset[10.17909/4gpn-y657]{http://dx.doi.org/10.17909/4gpn-y657}.
The data were obtained from the Mikulski Archive for Space Telescopes at the Space Telescope Science Institute, which is operated by the Association of Universities for Research in Astronomy, Inc., under NASA contract NAS 5-03127 for JWST. The observations are associated with JWST GO Cycle 1 and 2 programs 1282, 1549, 1584, 1640, 2025, 2260, 3228.

Part of this research was carried out at the Jet Propulsion Laboratory, California Institute of Technology, under a contract with the National Aeronautics and Space Administration (80NM0018D0004). The authors acknowledge support from NASA/Space Telescope Science Institute grants: JWST-GO-01640, JWST-GO-01584, and JWST-GO-01549. S.K. and T.K. acknowledge support from STFC Grant ST/Y002415/1. B.T. acknowledges support by the Programme National PCMI of CNRS/INSU with INC/INP cofunded by CEA and CNES.

\facilities{JWST}

\software{
Matplotlib \citep{matplotlib}, NumPy \citep{numpy}, SciPy \citep{scipy}, Seaborn \citep{seaborn}, Astropy \citep{astropy:2013, astropy:2018, astropy:2022}, LMFIT \citep{lmfit}, iSLAT \citep{iSLAT,iSLAT_code}.
}

\newpage
\appendix

\section{Additional figures for the entire sample} \label{app: sample_figures}
\begin{figure*}
\centering
\includegraphics[width=0.45\textwidth]{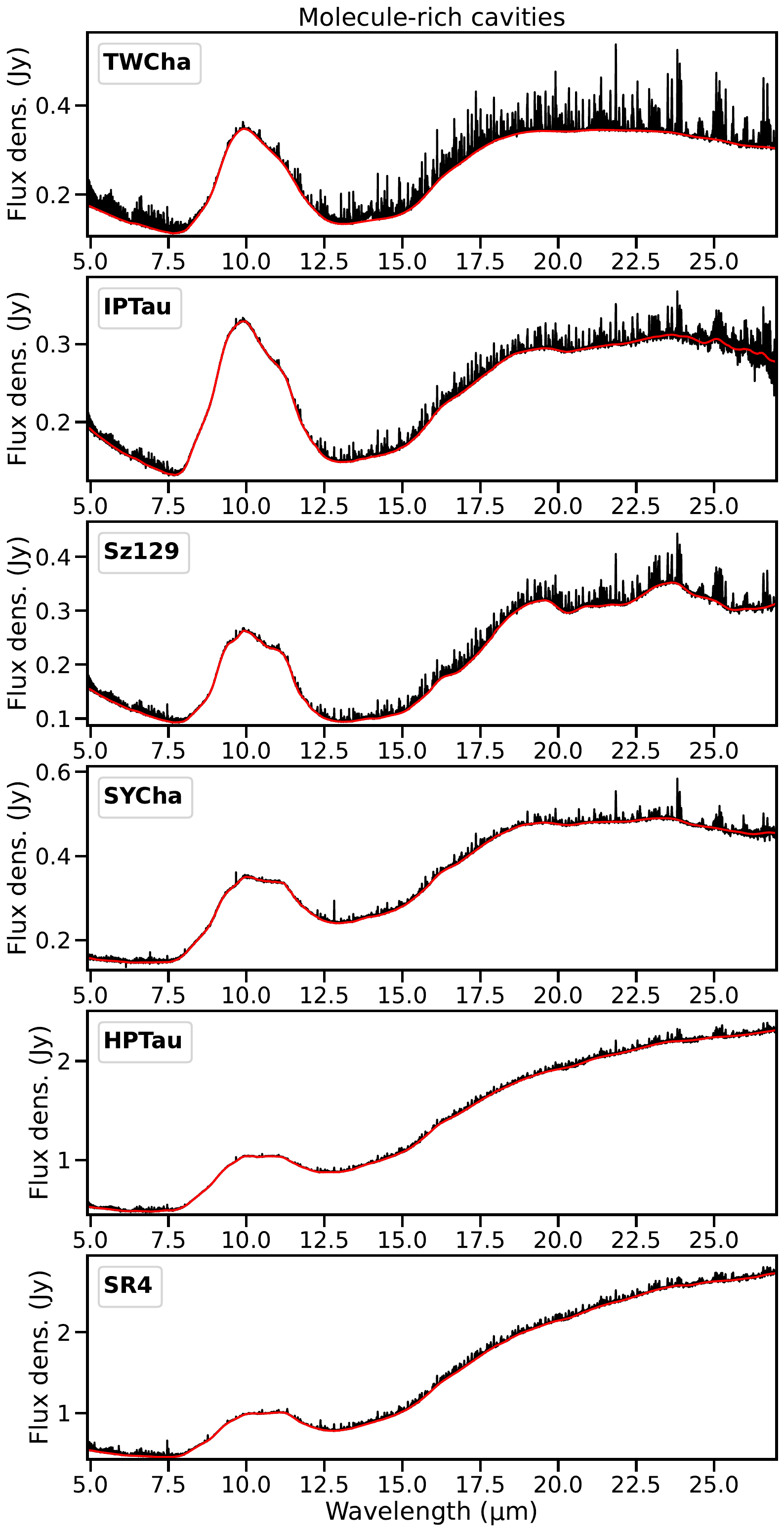} 
\includegraphics[width=0.45\textwidth]{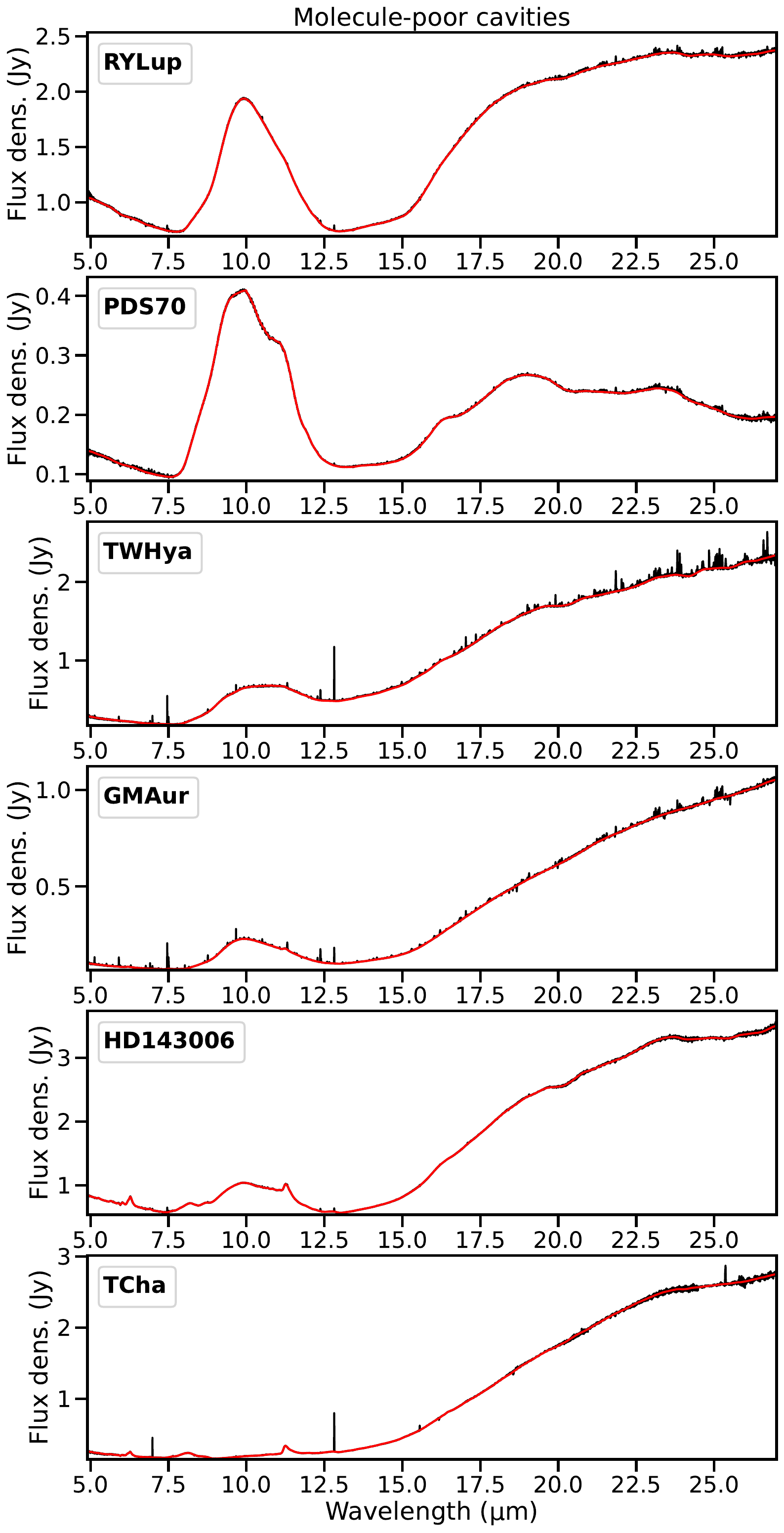} 
\caption{MIRI-MRS spectra for the cavity disks included in this work, with estimated continuum for each one marked in red. The MR/MP dichotomy does not seem to be related to different dust properties as visible in the 10~$\mu$m feature shape or prominence. }
\label{fig: spectra_original}
\end{figure*}

\begin{figure*}
\centering
\includegraphics[width=1\textwidth]{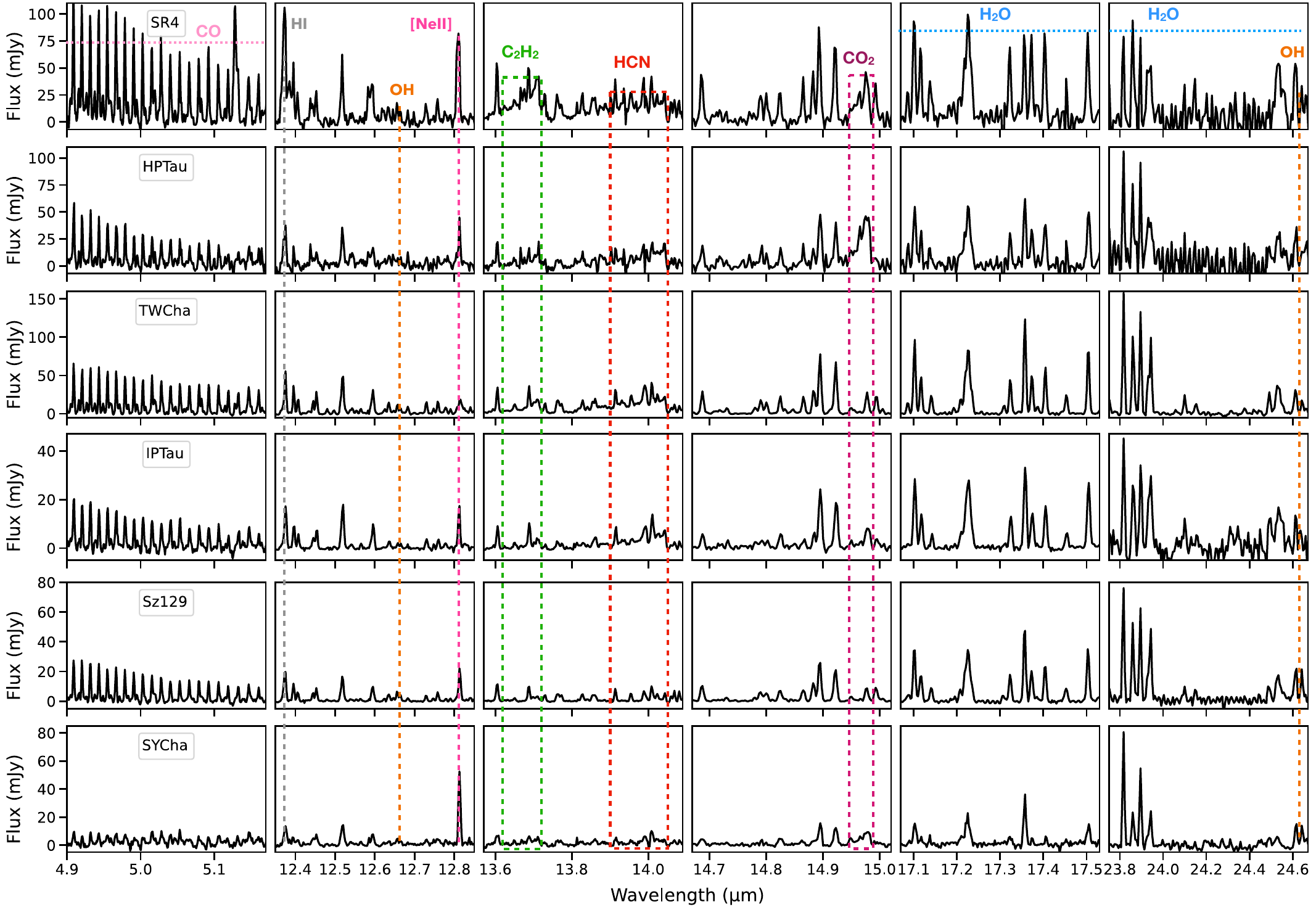} 
\caption{Continuum-subtracted MIRI spectra of MR cavities, showing prominent emission from all the typical molecules and atoms detected at infrared wavelengths in T~Tauri disks. Organic emission tends to be sub-luminous, as shown in Figure \ref{fig: lum_correl}.}
\label{fig: spectra_molecules_MR}
\end{figure*}

\begin{figure*}
\centering
\includegraphics[width=1\textwidth]{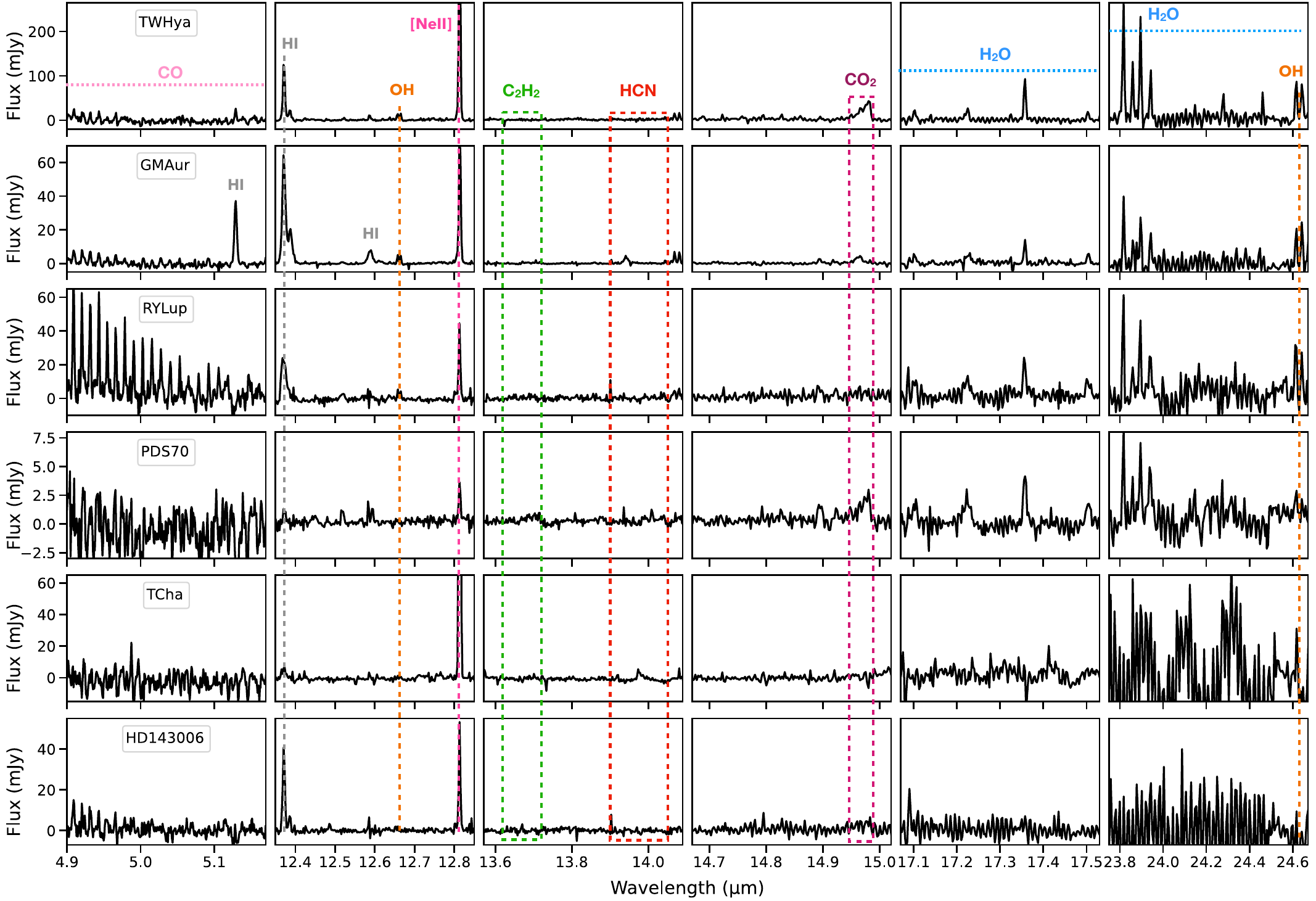} 
\caption{Continuum-subtracted MIRI spectra of MP cavities, showing prominent emission from HI and [NeII] and residual emission from CO, cold water, \ch{CO2}, and OH (not detected in T~Cha and HD~143006).}
\label{fig: spectra_molecules_MP}
\end{figure*}

Figure \ref{fig: spectra_original} shows the complete MIRI-MRS spectra of the cavity disks included in this work, with continuum marked in red for reference (this is the continuum that is subtracted before the analysis of gas emission). Figures \ref{fig: spectra_molecules_MR} and \ref{fig: spectra_molecules_MP} show portions of the continuum-subtracted spectra separated into MR and MP to illustrate their different molecular emission as described and discussed above in this work.

\section{The infrared index as a tracer of a dust cavity} \label{app: IRindex}
With the first studies of inner dust cavities being based on spatially-unresolved SED observations and the detection of a deficit at NIR wavelengths followed by an upturn in the far-IR, the slope of the SED at near- and mid-IR wavelengths has long been used as a proxy to detect the depletion of small dust grains in inner disks (Section \ref{sec: intro}). We briefly summarize in this section how different infrared indices have been used in the past and compare them to the new $n_{13-26}$ index measured from MIRI. We remark, however, that the introduction of these indices based on SED modeling explorations still dates back to works preceding the revolution provided by spatially-resolved millimeter interferometry imaging \citep[see references in Section \ref{sec: intro} and review by][]{vanderMarel23}. Due to the different radial distribution of $\mu$m dust vs mm grains, SED-based identifications of dust cavities pre-ALMA missed several (large) cavities that are now spatially resolved and known (Section \ref{sec: intro}). An updated modeling exploration of different IR indices in the context of different dust evolution effects (growth, settling, pebble drift, dust filtration through the mm cavity, inner rim structures, and especially the relative depletion of smaller vs larger grains) is long due and should be done in future work.

\begin{figure*}
\centering
\includegraphics[width=1\textwidth]{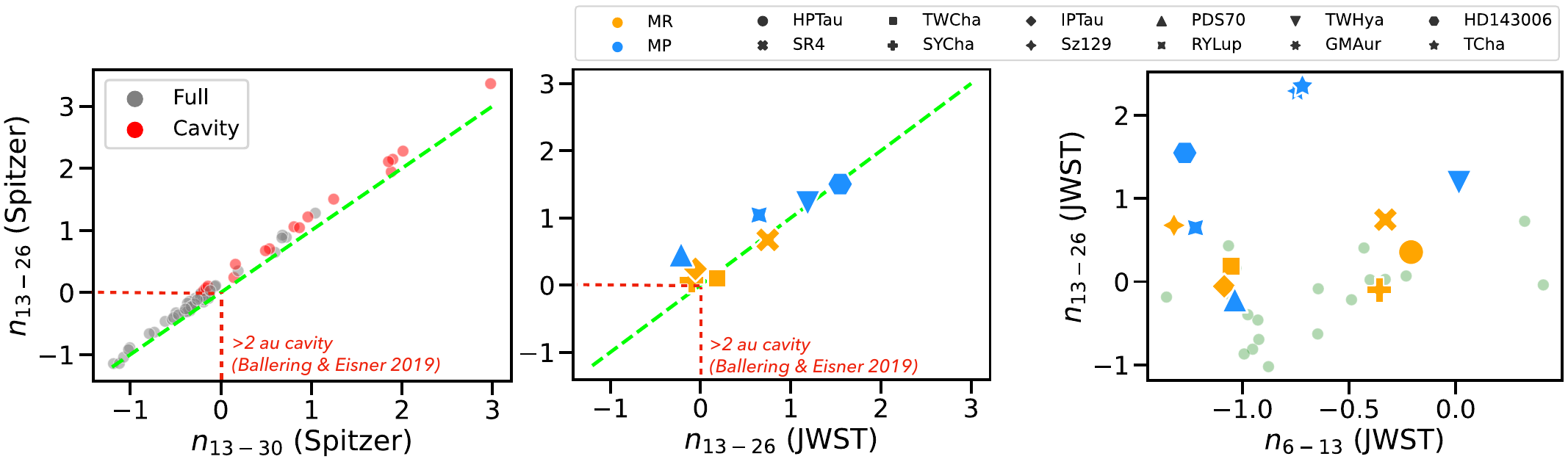} 
\caption{Comparison between different IR index definitions in Spitzer and JWST spectra and detection limits for dust cavities. The green dashed line shows the 1:1 relation. Left: Spitzer sample from Figure 9 in \cite{banz20}, with spatially-resolved millimeter continuum cavities marked in red. Some disks marked as ``full" may have spatially-unresolved cavities. For reference, disk models with indices $\gtrsim 0$ correspond to inner cavities of size $\gtrsim 2$~au in \cite{Ballering19}. Middle: sample from this work, where in common to the sample in \cite{banz20}. Right: MIRI-based $n_{13-26}$ index vs $n_{6-13}$ index for comparison to Figure 11 in \cite{furlan06}. A potential bifurcation unrelated to the MR/MP dichotomy (which are mixed in this plot) may be emerging in the sample in this plot and should be tested with larger samples in future work.}
\label{fig: IRindex_comparison}
\end{figure*}

While multiple spectral indices at wavelengths covered by Spitzer (5--37~$\mu$m) were explored, the $n_{13-30}$ ended up becoming the most used to detect an inner dust cavity \citep{furlan06,furlan09,furlan11,brown07}. These previous works determined that radially continuous, optically thick disk structures should have at most $n_{13-30} \lesssim 1$ in typical ISM dust/gas ratio (DGR) conditions (1:100) and found that most disks have instead $n_{13-30} \lesssim 0$, which they proposed to be ubiquitous evidence for dust settling in disk upper layers reducing DGR by factors of 10 to 1000 compared to ISM \citep{DAlessio2006}. Based on that, these works adopted a conservative limit of $n_{13-30} > 1$ to identify large dust cavities, with support from the first spatially-resolved images from millimeter interferometers where large dust cavities indeed had $n_{13-30} \gtrsim 1$. Multiple degeneracies affected SED modeling and a number of ``outliers" could not be explained by dust settling, down to $n_{13-30} \sim 0$ and in some cases even below \citep{furlan06,furlan09,DAlessio2006}. It should be noted that in these first studies, $0 \lesssim n_{13-30} \lesssim 1$ could only be reproduced with no or little dust settling, implying close to ISM values for the DGR that have later been argued against based on the excitation of IR molecular spectra \citep[e.g.][]{meijerink09,Greenwood19}. With more SMA and ALMA observations several millimeter cavities now demonstrate that the $n_{13-30}$ index can indeed be as small as 0, and sometimes lower, even in the case of very large mm cavities and even though it does not correlate simply with the size of the cavity at millimeter wavelengths \citep[Appendix D in][]{banz20}. 

However, the index is also affected by other parameters including the disk inclination angle along the line of sight \citep[e.g.][]{DAlessio2006,Ballering19,vanderMarel22}, which at high viewing angles ($\gtrsim 70$) can obscure the inner hot disk region and mimick a dust cavity in the SED. For this reason, in this work three disks that do not have a dust cavity detected from millimeter dust continuum with ALMA but have a positive IR index are excluded from the cavity sample due to their high inclination of $\sim 70$~deg, and they are included in the reference JDISCS-C1 sample used in this work (DoAr~25, MY~Lup, and IRAS~04385-2550). One of them, MY~Lup, has recently been presented and discussed in \cite{Salyk25} in the context of its peculiar molecular emission as possibly due to its high viewing angle or potentially an unresolved small inner cavity.
With multiple degeneracies and the possibility of different radial segregation of smaller versus larger dust grains (Section \ref{sec: intro}), it is expected that a clear cut in $n_{13-26}$ values may not be found to detect an inner dust cavity just from the SED. Based on the index values measured in spatially-resolved millimeter cavities, we have decided to adopt a limit of $n_{13-26} \gtrsim 0$ to identify a IR cavity from the SED \citep{banz20}. IP~Tau is an example of a case at the boundary, and we indicate its cavity as mm(+IR) in Table \ref{tab: sample} to reflect this uncertainty. Given the bifurcation discovered in Figure \ref{fig: sample_props}, it seems necessary in future work to explore and determine a more suitable way to classify IR cavities based on their SED.

Figure \ref{fig: IRindex_comparison} illustrates how the MIRI-based IR index $n_{13-26}$, measured at 13.095--13.113~$\mu$m and 26.3--26.4~$\mu$m where molecular emission is minimal as determined in \citet{banz25}, compares to the previously used $n_{13-30}$ from Spitzer. The two indices differ only slightly (with $n_{13-26}$ larger than $n_{13-30}$) when measured in the same spectrum (left plot in Figure \ref{fig: IRindex_comparison}), showing that spatially-resolved millimeter dust continuum cavities are still indicated by positive values of the $n_{13-26}$, as it was for $n_{13-30}$. The $n_{13-25}$ index in \citet{furlan06,furlan09} should be comparable to the new MIRI-based $n_{13-26}$ \citep[we do not use the range at 25~$\mu$m due to a cluster of strong water and OH lines, see e.g. ][]{banz25}. We include also the $n_{6-13}$ index used in \cite{furlan06}, which in MIRI we measure at 6.12--6.14~$\mu$m where there is a gap in molecular emission \citep{banz25}.

When comparing the $n_{13-26}$ index value as measured from MIRI to that previously measured in Spitzer spectra for the disks where both are available (middle plot in Figure \ref{fig: IRindex_comparison}), there is some evidence for variability in the SED, especially in the case of RY~Lup and PDS~70 \citep[the latter already reported in][]{perotti23}. We remark that the updated flux calibration in JDISCS reduction 9.0 has significantly changed all the $n_{13-26}$ index values previously measured in reduction version 8.0 and published in \citet{Arulanantham25}; in this work we have used the new values and we include those measured in the JDISCS-C1 sample in Table \ref{tab: JDISCS measurements}.

\begin{figure*}
\centering
\includegraphics[width=1\textwidth]{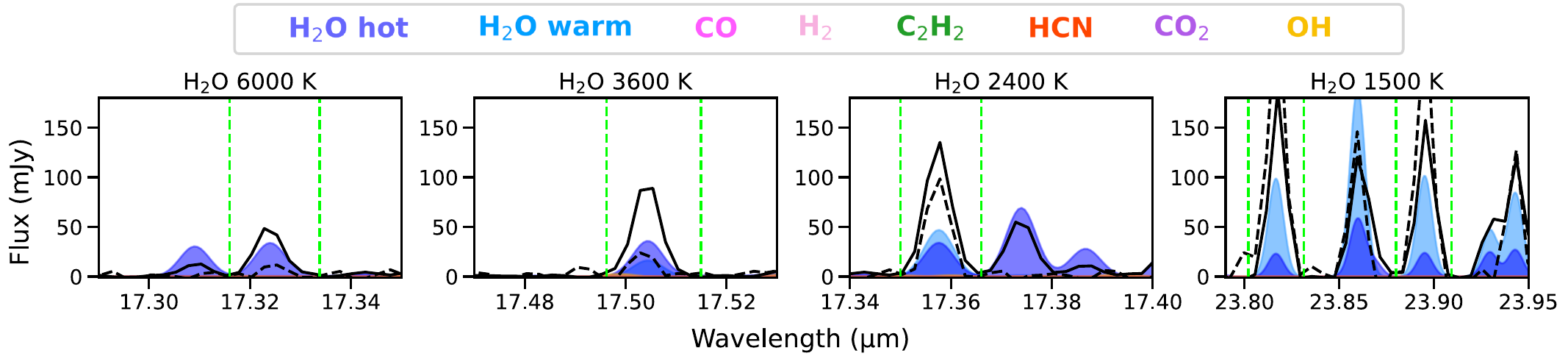} 
\includegraphics[width=1\textwidth]{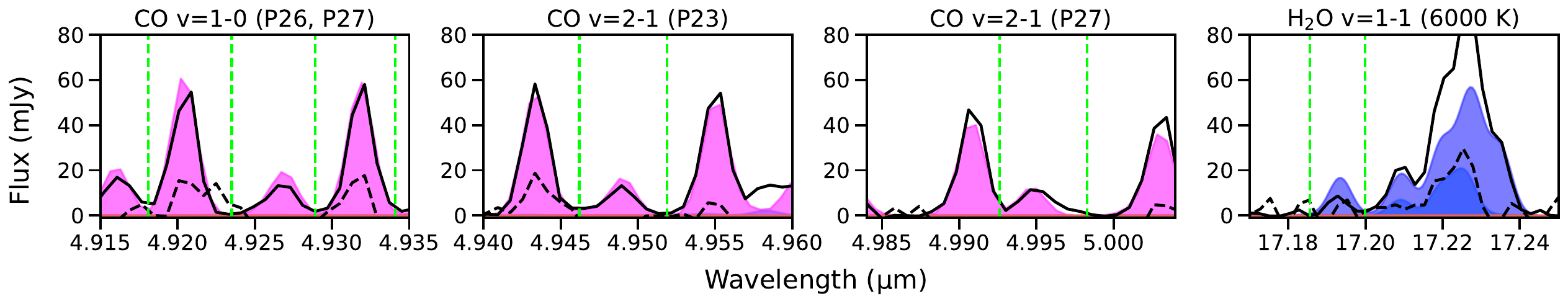} 
\includegraphics[width=1\textwidth]{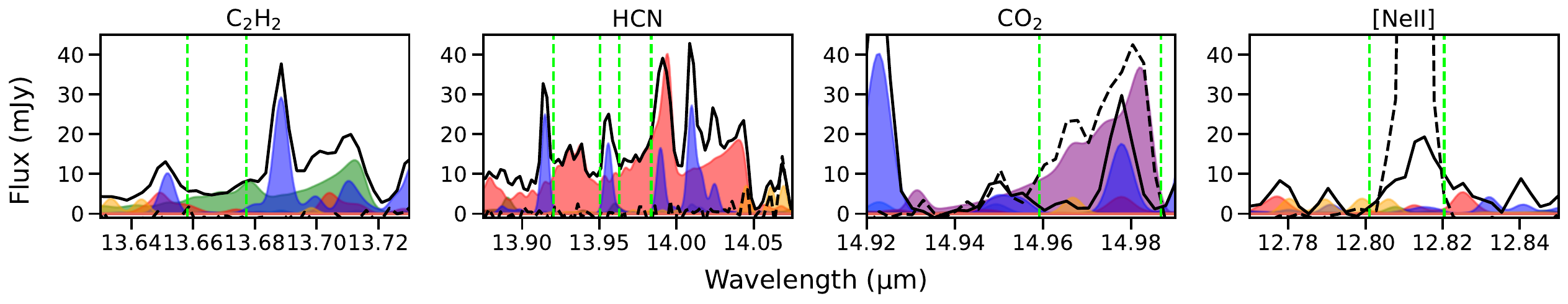} 
\includegraphics[width=1\textwidth]{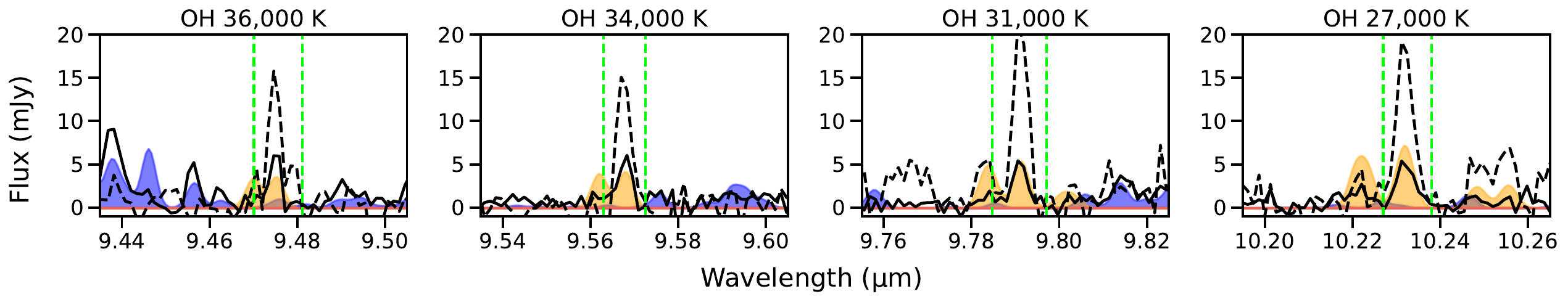} 
\includegraphics[width=1\textwidth]{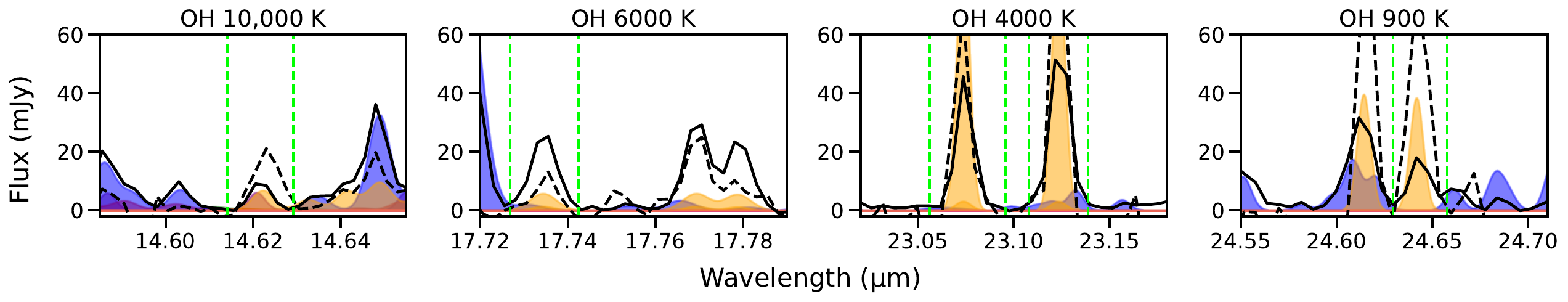} 
\caption{Spectral ranges used for line luminosity measurements in this work (see Table \ref{tab: lin_fit_params}), based on \cite{banz25} to determine lines that are free from blending (see Section \ref{app: new_lines_4_corr}). The MIRI spectra of TW~Cha (solid line) and TW~Hya (dashed line) are shown as examples of a molecule-rich (MR) and molecule-poor (MP) disk cavity respectively. Representative slab models from iSLAT \citep{iSLAT} are shown in each plot for comparison to the data. Vertical dashed lines show the range used to integrate the line flux in each case.}
\label{fig: line_ranges}
\end{figure*}

\section{Spectral ranges and molecular measurements} \label{app: new_lines_4_corr}
Figure \ref{fig: line_ranges} shows all the spectral ranges used to measure line luminosities in this work. The \ch{H2O}, CO, and some OH lines are directly taken from the line list provided in \cite{banz25}, available in iSLAT as ``MIRI\_general"\footnote{https://github.com/spexod/iSLAT}. This line list was carefully selected by accounting for emission from all the typical molecules observed with MIRI in disks, to minimize the contamination of each tracer from other species. Using the same contamination-minimization procedure, in this work we expand that line list with additional OH lines and we include emission from organics too. 
The general strategy for all molecules is to use uncontaminated lines that can be directly measured in the MIRI spectra without being impacted by the uncertainties and degeneracies of molecular slab modeling. All molecular spectra and contamination are carefully checked with iSLAT \citep{iSLAT}, which enables to inspect the individual transitions producing the emission from any molecules using data from HITRAN \citep{hitran20}. We remark that, as in \cite{banz25}, we included also the \ch{H2O} HITEMP line list \citep{HITEMP10} that is more complete to check for contamination from high-energy \ch{H2O} lines, especially at $< 18$~$\mu$m. The line list used in this work should therefore be straightforward to measure and reproduce in any other samples in the future, except for \ch{CO2} where some correction is needed (see below).
For measuring emission line properties from the spectra we use iSLAT, which implements the least-square minimization code \texttt{lmfit} \citep{lmfit} to perform single-Gaussian fits and measure the line flux and uncertainty for individual transitions and simply integrates the flux over a specific range in the case of the blended emission from the organics. 
Table \ref{tab:detections} reports detections for all these gas tracers in this sample, Tables \ref{tab: flux measurements} and \ref{tab: JDISCS measurements} report the line flux measurements and errors, and Table \ref{tab: lin_fit_params} reports linear regression results for Figures \ref{fig: lum_correl} and \ref{fig: lum_correl_IRindex}, using the \texttt{linregress} function in scipy \citep{scipy}.

For OH, we take lines across energy levels with careful consideration for blending with other molecules, especially \ch{H2O}. All the lines we select are free from contamination except for the 10,000~K line near 14.62~$\mu$m, which is contaminated by \ch{CO2} only in cases of strong emission from this molecule \citep[which are rare in the sample included in this work, see][]{Arulanantham25}. In the disk with strongest relative emission from \ch{CO2}, TW~Hya, this OH line is only contaminated by $\sim 20$\% of the measured line flux. For the 30,000~K OH lines, we select four and we sum their flux to increase S/N due to their typical weakness.
For the broad organic emission features, we select smaller ranges with minimal contamination from other molecules. We remark that to correctly consider contamination from water,  the HITRAN line list must be complemented with the HITEMP line list \citep{HITEMP10}, which includes several high-energy lines that significantly contaminate organic emission \citep{banz25}. In the case of \ch{C2H2}, the only range that can be used is around 13.66~$\mu$m, with mild contamination from HCN. The peak of \ch{C2H2} emission at 13.71--13.72~$\mu$m is instead contaminated by multiple water and HCN transitions (Figure \ref{fig: line_ranges}). In the case of HCN, there are two viable options that cover HCN emission dominated by slightly different upper energy levels; we tested correlations with both and decided to use the 13.97~$\mu$m because it better matches the energy range covered in \ch{C2H2} too (Table \ref{tab: lin_fit_params}). 

\ch{CO2} is the only case where avoiding contamination is impossible, because multiple species overlap with it including a strong \ch{H2O} line (which can be relatively easily removed), HCN, OH, and even HI (the 16-10 line). In this work, we decide to correct the measured \ch{CO2} line flux from HI contamination, by measuring and removing the line flux from the next HI line in the series (the 15-10 line at 16.41~$\mu$m), and from \ch{H2O} contamination by subtracting twice the flux measured in an uncontaminated line at 11.27~$\mu$m with same upper level energy ($\sim8200$~K) and lower Einstein-$A$ coefficient ($A_{ul}$) of $\sim 17 s^{-1}$ as the one contaminating \ch{CO2} (which has $\sim8200$~K and $A_{ul}$ of $\sim 62 s^{-1}$). The factor two in line flux is applied to account for the different $A_{ul}$, as determined in the case of TW~Cha where there is no \ch{CO2} emission. A little residual contamination from OH may still remain in \ch{CO2}, but we do not attempt to correct for that due to the complex excitation of OH (Section \ref{sec: OH}). We remark that a slab model to the \ch{CO2} emission would not necessarily give a more reliable estimate of the uncontaminated line flux, especially because slab model fits typically ignore the contamination from OH and HI and may not reproduce well the high-energy \ch{H2O} line either.

\section{Additional comparisons to previous work} \label{app: additional}

\begin{figure}
\centering
\includegraphics[width=0.45\textwidth]{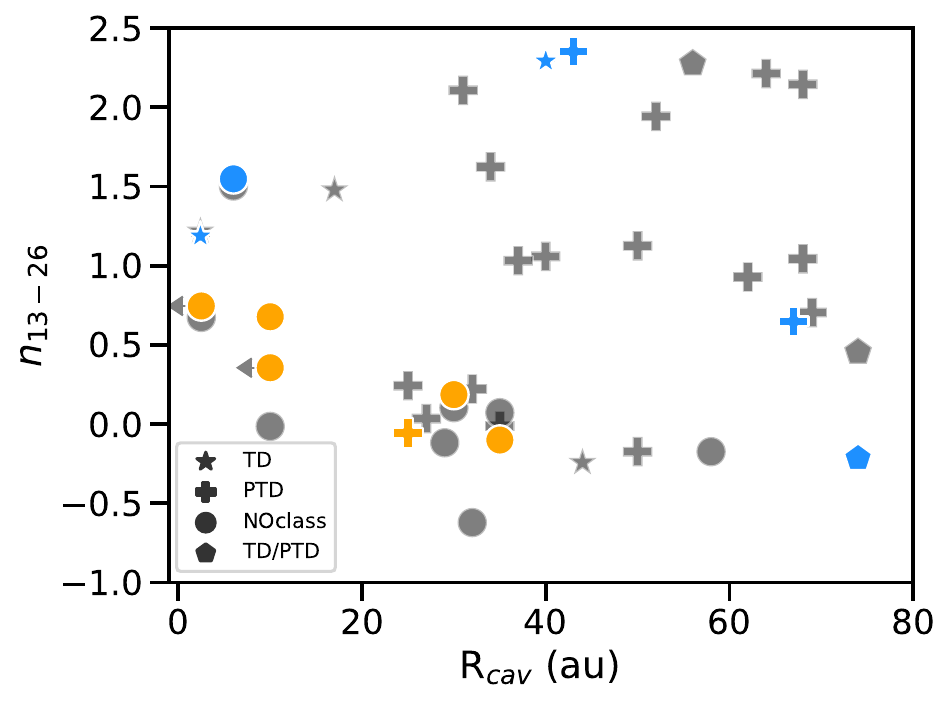} 
\caption{IR index $n_{13-26}$ against millimeter cavity radius R$_{cav}$, with TD and PTD classifications adopted from \cite{espaillat14,francis20}, which in a few cases do not agree (marked as TD/PTD in this figure). Some disks do not have a previous classification in these papers (marked as NOclass), including most of the MR cavities. The grey datapoints show millimeter cavities from \cite{pinilla18,francis20} matched to the IR index measured from Spitzer spectra (Appendix \ref{app: IRindex}). The molecular dichotomy presented in this work is included with same colors as in Figure \ref{fig: sample_props}, with $n_{13-26}$ from MIRI (Table \ref{tab: sample}).}
\label{fig: cavities_TD-PTD}
\end{figure}

\subsection{Comparison to previous classifications}
We compare the MR/MP dichotomy introduced in this work to previous classifications of cavity disks.
One of the dust cavities classifications that has taken more ground in the literature is that into ``transitional" (TD) and ``pre-transitional" (PTD) disks, based on the IR excess indicating residual inner dust belts within a larger dust cavity in the latter \citep{espaillat07,espaillat14}. This classification was based on spatially-unresolved SED modeling and has later been put into question due to limitations and degeneracies of using the near-IR excess as an indication of an inner dust belt \citep{vanderMarel23,Kaeufer2023}. By comparison to previous work, we can conclude that the TD/PTD classification does not simply map into the MR/MP dichotomy, as shown in Figure \ref{fig: cavities_TD-PTD}: PTD disks cover most of the parameter space in $n_{13-26}$ and R$_{cav}$ without distinctions in either parameter, while the MR/MP dichotomy shows very specific segregation in this plot (Figure \ref{fig: sample_props}). One possibility is that the separation into MR and MP is larger in $n_{13-26}$ than in the IR excess at shorter wavelengths.

Additional classifications have been proposed based on other observables.
\cite{salyk09} proposed the existence of two disk classes based on CO detections and accretion rates: ``cleared" inner disks and ``partially depleted" inner disks (with stronger CO emission and higher accretion rate). This classification too is different from the dichotomy introduced in this work, because MR and MP span the same range in accretion luminosity.
Another classification was proposed by  \citet{Owen2012, Owen2016} based on millimeter luminosity and accretion rate, with ``mm-faint" and low-accreting disks being consistent with disk dispersal through photoevaporation while ``mm-bright" disks being challenging to explain both with photoevaporative winds and planets. Again, this classification is not connected to the MR/MP dichotomy, which are not separated by accretion rates.
We should also note that disks around intermediate-mass stars and Herbig Ae/Be stars, the latter being excluded from this work, have often affected and possibly biased all these previous classifications. 
Large dust cavities are found to be more frequent in disks around stars $> 1.5 M_{\odot}$ \citep{vanderMarel23}, they appear to have high accretion rates and might have more extreme conditions than disk cavities around solar-mass stars \citep{ercolano_pascucci2017, Owen2012}.

\begin{figure}
\centering
\includegraphics[width=0.45\textwidth]{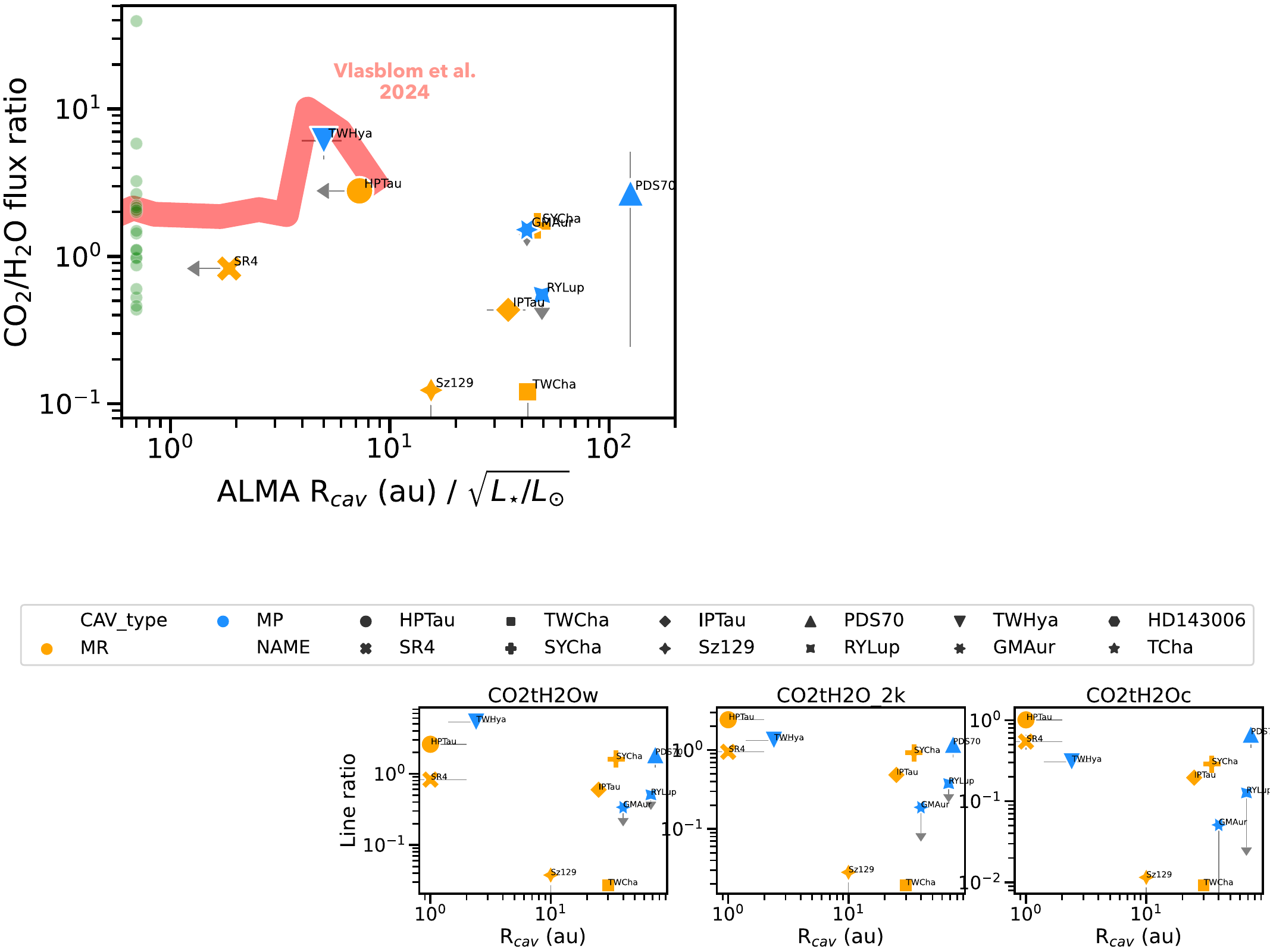} 
\caption{Ratio between \ch{CO2} and \ch{H2O} line fluxes as a function of cavity size, compared to predictions from \cite{Vlasblom24}. The water line used in this figure is the  3600~K line at 17.50~$\mu$m selected in \cite{banz25} and reported in Table \ref{tab: lin_fit_params}. The green points show the reference JDISCS-C1 sample of full disks as in previous figures, placed at an x-axis value of 0.7 just for illustration.}
\label{fig: CO2_cavity}
\end{figure}

\subsection{On \ch{CO2} emission as a tracer of inner disk cavities} \label{sec: co2}
Recent modeling work by \cite{Vlasblom24} proposed that the \ch{CO2} Q-branch emission relative to water emission observed with MIRI may be a tracer of small inner disk cavities, with larger \ch{CO2}/\ch{H2O} ratios when the cavity size is in between the \ch{H2O} and \ch{CO2} snowlines (between 0.5 and 2~au in their fiducial model). We test this idea in Figure \ref{fig: CO2_cavity}, where the \ch{CO2}/\ch{H2O} ratio measured in this sample is compared to model predictions from Figure B.7 in \cite{Vlasblom24} as normalized to the square root of the stellar luminosity. In this work, we use the 3600~K water line at 17.50~$\mu$m selected in \cite{banz25}, which is sensitive to the warm water reservoir; the model track shown in the figure is also updated to this line. The models show a peak in the \ch{CO2}/\ch{H2O} flux ratio at values of $\sim 4$~au / $\sqrt{L_{\star}/L_{\odot}}$, decreasing at larger radii when the cavity size reaches and surpasses the \ch{CO2} snowline. 

In this work, we only have 3 disks that overlap with model predictions from \cite{Vlasblom24}: TW~Hya, HP~Tau and SR~4; the latter two, however, have an uncertain inner disk cavity (positive IR index but undetected mm cavity). The model peak in the \ch{CO2}/\ch{H2O} ratio overlaps with TW~Hya, while the rest of the sample sits at values $> 10$~au / $\sqrt{L_{\star}/L_{\odot}}$ and shows a large range of ratios (as small as $< 0.1$ and as large as 2). On the other hand, the reference sample of full disks also span a similar range of \ch{CO2}/\ch{H2O} ratios (green points in the figure), showing that either small dust cavities are very common, or that this ratio might be sensitive to other processes than just an inner dust cavity. It is clear that more disks with small cavities are needed to test the models and any dependence of the \ch{CO2}/\ch{H2O} ratio on cavity size. Moreover, if the mm cavity size does not trace the innermost cavity size in smaller dust, the comparison to the models (which assume a completely devoid inner cavity for simplicity) needs to be updated with more sophisticated models that account for a different distribution of sub-$\mu$m dust and larger solid grains.

\subsection{The peculiar case of PDS~70} \label{sec: pds70}
The water emission observed in PDS~70 stands out as an outlier in the sample of disks cavities under multiple aspects, as shown above in Figures \ref{fig: lum_correl_IRindex} and \ref{fig: water_diagrams}. Its spectrum is consistent with a single, warm (T$\sim 400$~K), optically thick ($\sim 5 \times 10^{18}$~cm$^{-2}$) inner water reservoir emitting from $\sim 0.1$~au (assuming the resolved line broadening is from Keplerian rotation around the star) unlike any other MP cavity, where emission is more optically thin and comes from larger radii. It is the only disk so far where a single slab model to the emission of rotational lines at 10--27~$\mu$m does not over-predict the ro-vibrational water band at 6--8~$\mu$m \citep{perotti23}, again pointing to a small and highly dense emitting region. The emitting area is small and of the order of $\sim 0.007$~au$^2$. The origin of this unusual emission could be an inner dust belt where water survives, which may be detected in millimeter emission with ALMA \citep{francis20,benisty21} and could still be fed by some residual filtration of ice carried inward on small dust grains even through the very large cavity \citep{pinilla24,Jang2024}. An alternative, interesting scenario that might offer an explanation for the peculiar water emission observed in this disk is that it could instead be from a circum-planetary disk around one of the protoplanets detected in this system \citep{benisty21}. The circum-planetary disk is not spatially resolved with ALMA, but the emitting area of water emission is small enough to be consistent with it.

The MIRI spectrum also offers a tentative detection of HCN and \ch{C2H2} too, which seem to be colder than in other disks (Perotti et al., in prep.).
The IR index has been observed to be highly variable between -0.22 in the MIRI spectrum and 0.5 in a previous Spitzer spectrum in 2007 \citep[][]{perotti23}, showing highly variable inner disk dust content or geometry \citep{gaidos24,Jang2024}. This system should be monitored with MIRI to determine if the molecular spectrum is variable too. The SED variability in PDS~70, recently observed in another MP cavity \citep[T~Cha,][]{Xie_2025}, also poses the question whether it could be common in MP cavities in general, something that should be tested in future work.

\begin{deluxetable*}{l c c c c c | c | c}
\tablecaption{\label{tab: lin_fit_params} Linear regression parameters for correlations with log$L_{\rm{acc}}$ as measured in the reference JDISCS-C1 sample excluding disks with cavities.}
\tablehead{Line ID & $E_u$ [K] & $\lambda$ [$\mu$m] & a $ (\sigma_a)$ & b $ (\sigma_b)$ & PCC & PCC (log$L_{\rm{\star}}$) & PCC ($n_{13-26}$)}
\startdata
CO $v$=1-0 & 4862, 5004 & 4.92, 4.93 & 0.594 (0.066) & -4.235 (0.095) & 0.91 & 0.52 & -0.43 \\
CO $v$=2-1 & 7501, 8031 & 4.95, 4.99 & 0.554 (0.053) & -4.582 (0.060) & 0.92 & 0.60 & -0.55 \\
\ch{H2O} $v$=1-1 & 6006 & 17.19 & 0.468 (0.069) & -5.718 (0.078) & 0.88 & 0.45 & -0.42 \\
\ch{H2O} 6000 K & 6052 & 17.32 & 0.523 (0.072) & -5.087 (0.097) & 0.88 & 0.59 & -0.32 \\
\ch{H2O} 3600 K & 3646 & 17.50 & 0.533 (0.078) & -4.787 (0.114) & 0.86 & 0.54 & -0.37 \\
\ch{H2O} 2400 K & 2433 & 17.36 & 0.488 (0.091) & -4.706 (0.132) & 0.79 & 0.56 & -0.31 \\
\ch{H2O} 1500 K & 1448, 1615 & 23.82, 23.90 & 0.446 (0.091) & -4.436 (0.132) & 0.77 & 0.53 & -0.28 \\
OH 30,000 K & 26716--35575 & 9.48--10.23 & 0.333 (0.074) & -5.323 (0.095) & 0.81 & 0.74 & -0.52 \\
OH 10,000 K & 10754 & 14.62 & 0.436 (0.079) & -5.486 (0.115) & 0.80 & 0.48 & -0.38 \\
OH 6000 K & 5495, 7072 & 17.73, 20.00 & 0.474 (0.061) & -5.287 (0.084) & 0.89 & 0.58 & -0.37 \\
OH 4000 K & 3473, 4104 & 23.07, 25.09 & 0.418 (0.052) & -5.136 (0.075) & 0.89 & 0.56 & -0.46 \\
OH 900 K & 875 & 24.64 & 0.203 (0.082) & -5.741 (0.107) & 0.58 & 0.32 & -0.01 \\
HCN & 2000--3700 & 13.963--13.984  & 0.395 (0.108) & -4.945 (0.160) & 0.68 & 0.61 & -0.28 \\
\ch{C2H2} & 1800--3500 & 13.658--13.677 & 0.284 (0.111) & -5.086 (0.172) & 0.58 & 0.61 & -0.47 \\
\ch{CO2} & 1000--2800 & 14.959--14.987 & 0.302 (0.119) & -4.813 (0.177) & 0.54 & 0.38 & -0.01 \\
\enddata
\tablecomments{Linear relations are in the form $\rm{log}(L_{line}/L_\odot) = a \times \rm{log}(L_{acc}/L_\odot) + b$. The second and third columns indicate the value in upper level energy and wavelength included in each tracer, or their range if multiple lines are included (See Appendix \ref{app: new_lines_4_corr}). In the case of organics, the observed emission features are blends of 10--30 lines even over the small ranges considered here. The PCC value is the Pearson linear correlation coefficient; the second to last column reports the PCC with log$L_{\rm{\star}}$, showing that it is systematically lower than that measured with log$L_{\rm{acc}}$ except for the case of \ch{C2H2}.}
\end{deluxetable*}

\begin{deluxetable*}{l c c c c c c c c c c}
\tabletypesize{\small}
\tablewidth{0pt}
\tablecaption{\label{tab: flux measurements} Line flux measurements from MIRI spectra of disk cavities included in this work.}
\tablehead{Target & \ch{H2O} 6000 K & \ch{H2O} 3600 K & \ch{H2O} 2400 K & \ch{H2O} 1500 K & \ch{H2O} 1448 K & \ch{H2O} $v$=1-1 & CO $v$=1-0 & CO $v$=2-1 }
\tablecolumns{9}
\startdata
GMAur & 0.13(0.02) & 0.46(0.05) & 0.79(0.04) & 3.92(0.76) & 2.33(0.42) & 0.04(0.04) & 2.75(0.47) & -0.36(0.23) \\
HD143006 & 0.21(0.26) & -0.03(0.29) & 0.28(0.23) & -0.98(1.94) & -1.11(0.99) & 0.15(0.20) & 2.95(0.69) & -0.80(1.51) \\
HPTau & 2.21(0.23) & 3.96(0.15) & 4.41(0.35) & 12.86(1.03) & 6.97(0.65) & 0.14(0.18) & 12.01(1.28) & 1.59(0.58) \\
IPTau & 1.15(0.09) & 2.36(0.04) & 2.91(0.10) & 6.11(0.55) & 3.47(0.32) & 0.07(0.03) & 5.74(0.49) & 0.38(0.30) \\
PDS70 & 0.11(0.05) & 0.20(0.06) & 0.40(0.09) & 1.06(0.09) & 0.58(0.04) & 0.01(0.04) & 0.30(0.92) & -0.60(0.74) \\
RYLup & 0.36(0.27) & 1.19(0.40) & 1.67(0.12) & 7.17(1.15) & 3.62(0.53) & 0.17(0.18) & 15.26(1.12) & 1.30(3.17) \\
SR4 & 4.98(0.29) & 6.25(0.25) & 5.66(0.32) & 8.88(1.98) & 4.57(0.79) & 0.84(0.15) & 27.73(1.47) & 6.19(0.72) \\
SYCha & 0.58(0.15) & 1.65(0.08) & 2.92(0.07) & 9.39(0.37) & 5.45(0.22) & 0.22(0.14) & 2.17(0.29) & 0.67(0.69) \\
Sz129 & 1.44(0.04) & 2.64(0.04) & 3.52(0.04) & 9.72(0.26) & 5.38(0.11) & 0.30(0.13) & 7.15(0.45) & 0.99(0.22) \\
TCha & 0.76(0.36) & 0.54(0.26) & -0.28(0.30) & -0.97(2.36) & -1.71(1.18) & 0.33(0.15) & -0.21(1.93) & -0.34(1.29) \\
TWCha & 3.43(0.17) & 6.27(0.16) & 8.64(0.17) & 19.45(0.48) & 10.41(0.30) & 0.56(0.06) & 14.25(1.07) & 3.93(0.44) \\
TWHya & 0.69(0.15) & 1.42(0.28) & 5.71(0.35) & 29.87(1.71) & 16.06(0.82) & 0.25(0.26) & 4.91(2.02) & -2.95(1.32) \\
\hline
\hline
Target & OH 30,0000 K & OH 10,0000 K & OH 6000 K & OH 4000 K & OH 900 K & HCN & \ch{C2H2} & \ch{CO2} \\
\hline
GMAur & 3.44(0.31) & 0.16(0.04) & 0.94(0.34) & 3.12(0.19) & 1.68(0.10) & 0.05(0.05) & 0.05(0.04) & -0.69(0.14) \\
HD143006 & -1.17(1.74) & -0.04(0.13) & 0.14(0.49) & 0.07(0.83) & -0.07(1.37) & -0.21(0.15) & -0.15(0.14) & 0.58(1.42) \\
HPTau & 3.63(0.81) & 1.15(0.11) & 2.95(0.77) & 5.40(0.36) & 1.21(0.20) & 1.62(0.25) & 0.73(0.22) & 11.01(0.45) \\
IPTau & 1.26(0.34) & 0.31(0.08) & 0.98(0.06) & 1.93(0.32) & 0.53(0.36) & 1.14(0.06) & 0.27(0.06) & 1.03(0.15) \\
PDS70 & -1.01(1.04) & 0.07(0.02) & 0.07(0.05) & 0.50(0.13) & 0.10(0.03) & 0.05(0.05) & 0.11(0.07) & 0.55(0.47) \\
RYLup & 2.13(2.57) & -0.10(0.14) & 1.28(0.42) & 3.71(0.29) & 1.72(0.45) & 0.00(0.23) & 0.09(0.32) & 0.65(0.78) \\
SR4 & 4.88(1.25) & 1.48(0.21) & 4.96(0.96) & 5.68(0.33) & 1.87(0.56) & 4.31(0.28) & 5.93(0.35) & 5.17(0.49) \\
SYCha & 1.22(0.84) & 0.45(0.08) & 0.76(0.13) & 1.66(0.22) & 0.96(0.03) & 0.79(0.27) & 0.38(0.21) & 2.66(0.38) \\
Sz129 & 2.66(0.32) & 0.60(0.04) & 1.76(0.20) & 3.60(0.20) & 1.25(0.04) & 0.14(0.03) & 0.01(0.06) & 0.33(0.11) \\
TCha & 1.08(1.00) & -0.10(0.10) & 0.58(0.90) & 0.82(1.00) & -0.81(1.11) & 0.63(0.22) & -0.22(0.15) & -0.28(0.88) \\
TWCha & 3.01(0.49) & 0.98(0.11) & 2.36(0.09) & 4.05(0.16) & 1.23(0.14) & 4.42(0.07) & 1.58(0.07) & 0.76(0.15) \\
TWHya & 9.25(1.03) & 0.74(0.21) & 1.49(0.15) & 8.38(1.19) & 5.84(0.33) & -0.19(0.39) & -0.29(0.56) & 8.65(1.34) \\
\enddata
\tablecomments{Lines are labeled as defined in Table \ref{tab: lin_fit_params} and Section \ref{sec: analysis}. Line fluxes and errors are reported in units of $10^{-15}$ erg s$^{-1}$ cm$^{-2}$. 1-$\sigma$ errors are shown in parentheses. Negative flux values in non-detections are typically due to fringe residuals (see e.g. Figure \ref{fig: spectra_molecules_MP}). }
\end{deluxetable*}

\begin{deluxetable*}{l c c c c c c c c c c}
\tabletypesize{\small}
\tablewidth{0pt}
\tablecaption{\label{tab: JDISCS measurements} Line flux measurements from MIRI spectra of full disks included in this work.}
\tablehead{Target & \ch{H2O} 6000 K & \ch{H2O} 3600 K & \ch{H2O} 2400 K & \ch{H2O} 1500 K & \ch{H2O} 1448 K & \ch{H2O} $v$=1-1 & CO $v$=1-0 & CO $v$=2-1 }
\tablecolumns{9}
\startdata
AS205N & 30.55(2.04) & 70.07(2.71) & 99.27(2.66) & 162.87(10.62) & 88.58(5.11) & 9.99(1.49) & 169.35(12.88) & 97.99(12.36) \\
AS209 & 3.55(0.27) & 6.03(0.56) & 4.69(1.33) & 16.53(3.54) & 9.28(1.76) & 1.08(1.11) & 32.62(3.07) & 7.00(3.29) \\
CITau & 5.36(0.30) & 7.15(0.13) & 7.31(0.21) & 11.91(0.34) & 5.70(0.23) & 1.09(0.02) & 29.52(1.92) & 12.84(0.88) \\
DoAr25 & 1.21(0.09) & 2.52(0.13) & 2.80(0.12) & 4.71(0.53) & 2.27(0.34) & 0.21(0.03) & 7.28(0.98) & -0.11(0.57) \\
DoAr33 & 0.75(0.05) & 1.70(0.05) & 2.12(0.11) & 3.34(0.18) & 1.62(0.11) & 0.07(0.03) & 2.42(0.49) & -0.27(0.36) \\
Elias20 & 6.91(0.24) & 16.93(0.44) & 25.23(0.54) & 54.15(0.87) & 27.91(0.52) & 1.93(0.37) & 65.96(4.50) & 25.79(2.60) \\
Elias24 & 14.43(0.47) & 36.30(0.93) & 58.43(0.76) & 123.84(1.01) & 64.49(0.06) & 2.68(0.29) & 95.99(6.63) & 33.02(5.56) \\
Elias27 & 5.70(0.20) & 11.78(0.21) & 14.38(0.37) & 25.93(0.61) & 13.46(0.17) & 1.13(0.13) & 47.21(3.52) & 20.33(2.59) \\
FZTau & 15.34(1.00) & 29.10(1.30) & 35.43(0.61) & 51.04(0.82) & 24.99(0.24) & 4.04(0.20) & 104.67(11.29) & 58.17(7.74) \\
GKTau & 3.61(0.34) & 7.01(0.31) & 10.08(0.25) & 26.22(0.73) & 14.37(0.47) & 0.64(0.13) & 27.08(1.49) & 8.00(0.78) \\
GOTau & 0.16(0.02) & 0.46(0.02) & 0.52(0.02) & 1.12(0.15) & 0.59(0.07) & 0.01(0.01) & 1.51(0.14) & -0.13(0.14) \\
GQLup & 2.87(0.18) & 7.54(0.26) & 12.83(0.10) & 34.46(0.43) & 18.27(0.25) & 0.67(0.16) & 47.05(1.58) & 9.60(1.73) \\
IQTau & 3.32(0.16) & 4.11(0.17) & 4.91(0.19) & 11.36(0.41) & 6.30(0.23) & 0.95(0.02) & 7.95(1.65) & 8.35(1.52) \\
IRAS-04385 & 1.36(0.11) & 4.50(0.22) & 9.54(0.28) & 26.39(1.17) & 15.06(0.80) & 0.40(0.15) & 3.91(0.30) & 1.13(0.36) \\
MYLup & 0.03(0.07) & 0.23(0.03) & 0.61(0.03) & 1.64(0.34) & 0.81(0.17) & -0.09(0.06) & 1.06(0.32) & -0.66(0.27) \\
RULup & 15.35(1.02) & 26.28(1.24) & 31.23(0.60) & 44.25(2.46) & 22.40(1.35) & 3.40(0.55) & 103.17(8.97) & 42.74(6.27) \\
Sz114 & 1.24(0.06) & 3.14(0.03) & 4.49(0.04) & 11.64(0.50) & 6.44(0.17) & 0.37(0.08) & 10.28(0.44) & 2.88(1.04) \\
VZCha & 5.01(0.24) & 7.18(0.28) & 7.90(0.24) & 18.20(0.12) & 9.47(0.08) & 0.89(0.05) & 46.48(3.98) & 15.50(2.04) \\
WSB52 & 12.07(0.40) & 30.49(0.48) & 43.45(0.42) & 67.68(1.56) & 34.85(0.88) & 2.02(0.10) & 29.54(1.29) & 10.59(0.59) \\
\hline
\hline
$n_{13-26}$ & OH 30,0000 K & OH 10,0000 K & OH 6000 K & OH 4000 K & OH 900 K & HCN & \ch{C2H2} & \ch{CO2} \\
\hline
0.03 & 3.14(17.63) & 13.17(2.10) & 18.49(1.99) & 21.95(4.18) & 6.86(1.47) & 60.38(4.29) & 40.47(4.23) & 155.34(6.75) \\
-0.04 & -10.66(9.50) & 3.74(0.28) & 3.94(1.93) & 7.42(1.85) & 6.57(1.98) & 1.84(3.39) & 1.24(2.94) & 6.64(5.22) \\
-0.18 & 3.79(0.62) & 2.00(0.09) & 3.41(0.13) & 4.55(0.15) & 0.53(0.61) & 9.44(0.19) & 8.85(0.16) & 7.04(0.35) \\
0.43 & 1.14(0.30) & 0.16(0.05) & 0.44(0.09) & 1.52(0.07) & 0.25(0.21) & 8.60(0.09) & 2.07(0.10) & 5.18(0.19) \\
-0.87 & -0.09(0.52) & 0.44(0.06) & 0.50(0.07) & 1.37(0.12) & 0.11(0.16) & 4.49(0.10) & 5.02(0.10) & 4.50(0.24) \\
-0.69 & 4.77(0.69) & 3.90(0.75) & 5.85(0.26) & 8.65(0.30) & 0.98(0.33) & 24.71(0.43) & 32.26(0.40) & 18.75(0.53) \\
-0.63 & 3.99(3.51) & 7.45(1.08) & 7.62(1.21) & 11.85(0.33) & 4.62(2.07) & 50.60(1.81) & 50.62(1.97) & 54.00(1.46) \\
-0.46 & 2.64(0.52) & 2.39(0.11) & 3.39(0.15) & 5.10(0.16) & 2.34(0.71) & 2.17(0.14) & 0.30(0.21) & 10.24(0.54) \\
-0.81 & 13.33(3.02) & 6.29(0.57) & 7.07(0.49) & 10.08(0.74) & 2.14(0.32) & 13.99(0.37) & 5.13(0.37) & 41.61(0.91) \\
-0.09 & 5.38(1.20) & 2.24(0.23) & 4.18(0.18) & 7.59(0.44) & 3.14(0.61) & 0.60(0.18) & 0.17(0.19) & 3.23(0.42) \\
0.16 & 0.52(0.13) & 0.10(0.03) & 0.30(0.08) & 0.69(0.06) & 0.10(0.04) & 0.60(0.03) & 2.02(0.03) & 0.93(0.06) \\
-0.19 & 5.07(0.94) & 1.61(0.07) & 4.38(0.08) & 9.20(0.35) & 1.40(0.41) & 3.86(0.20) & 0.01(0.27) & 4.53(0.42) \\
-0.4 & 3.46(0.61) & 1.33(0.25) & 2.97(0.13) & 3.58(0.24) & 1.48(0.24) & 6.10(0.20) & 2.39(0.13) & 2.16(0.31) \\
0.73 & 1.70(0.52) & 0.75(0.11) & 0.88(0.27) & 1.27(0.27) & 1.12(0.53) & 4.17(0.15) & 3.19(0.15) & 14.56(0.29) \\
0.4 & 0.07(0.34) & 0.50(0.10) & 0.07(0.13) & 0.31(0.15) & 0.36(0.15) & 1.36(0.11) & 0.61(0.14) & 8.89(0.33) \\
0.07 & 4.68(2.90) & 4.86(0.82) & 7.20(1.68) & 11.98(0.76) & 2.91(0.06) & 16.68(0.79) & 2.25(0.83) & 25.60(1.97) \\
0.02 & 1.25(0.24) & 1.33(0.22) & 0.64(0.11) & 1.57(0.14) & 1.25(0.09) & 2.33(0.08) & 3.02(0.11) & 18.25(0.17) \\
-1.02 & 3.66(0.64) & 1.57(0.07) & 4.05(0.15) & 5.67(0.17) & 0.46(0.12) & 6.02(0.18) & 6.06(0.12) & 3.13(0.27) \\
-0.22 & 5.36(0.73) & 4.66(0.20) & 3.06(0.85) & 4.62(0.33) & 3.89(0.42) & 24.10(0.38) & 21.54(0.42) & 65.32(0.92) \\
\enddata
\tablecomments{Same as Table \ref{tab: flux measurements} but for the JDISCS-C1 sample. The $n_{13-26}$ index is included in the bottom half of the table (which corresponds to the same targets as in the upper half) because the improved absolute flux calibration in JDISCS reduction 9.0 has changed the continuum flux by up to 10--20\% in channel 4 and in turn it has changed all the measured values from those previously measured in reduction 8.0 as published in \citet{Arulanantham25}.}
\end{deluxetable*}

\bibliography{cavities_MIRI.bib}{}
\bibliographystyle{aasjournal}

\end{document}